\documentclass[12pt]{article}
\usepackage{subeqn}
\usepackage{epsfig}
\newcommand{\be}{\begin{equation}}
\newcommand{\ee}{\end{equation}}

\newcommand{\no}{\noindent}
\newcommand{\ce}{\begin{center}}
\newcommand{\nc}{\end{center}}

\makeatletter
\@addtoreset{equation}{section}
\makeatother

\def\sqr#1#2{{\vcenter{\vbox{\hrule height.#2pt
 \hbox{\vrule width.#2pt height#1pt \kern#1pt
 \vrule width.#2pt} \hrule height.#2pt}}}}

\def\operp{\hbox{${\kern+.25em{\bigcirc}
\kern-.85em\bot\kern+.85em\kern-.25em}$}}
\def\lsim{\;\raise0.3ex\hbox{$<$\kern-0.75em\raise-1.1ex\hbox{$\sim$}}\;}
\def\gsim{\;\raise0.3ex\hbox{$>$\kern-0.75em\raise-1.1ex\hbox{$\sim$}}\;}
\def\no{\noindent}

\def\ce{\centerline}
\def\ve{\vfill\eject}
\def\rdots{\mathinner{\mkern1mu\raise1pt\vbox{\kern7pt\hbox{.}}\mkern2mu
 \raise4pt\hbox{.}\mkern2mu\raise7pt\hbox{.}\mkern1mu}}

\def\e e{$e^+ e^-$ }

\begin{document}

\ce{\bf SOLITONIC MODELS BASED ON QUANTUM GROUPS} 
\ce{\bf AND THE STANDARD MODEL}

\vskip.3cm

\ce{\it Robert J. Finkelstein}
\vskip.3cm

\ce{Department of Physics and Astronomy}
\ce{University of California, Los Angeles, CA 90095-1547}

\vskip1.0cm

\no {\bf Abstract.}  The idea that the elementary particles might have
the symmetry of knots has had a long history.  In any current formulation of this idea, however, the knot must be quantized.  The
present review is a summary of a small set of papers that 
began as an attempt
to correlate the properties of quantized knots with the empirical
properties of the elementary particles.  As the ideas behind these
papers have developed over a number of years the model has evolved,
and this review is intended to present the model in its current form.
The original picture of an elementary fermion as a solitonic knot of
field, described by the trefoil representation of $SU_q(2)$, has 
expanded into its current form in which a knotted field is 
complementary to a composite structure composed of three or more preons
that in turn are described by the fundamental representation of
$SL_q(2)$.  These complementary descriptions may be interpreted as
describing single composite particles composed of three or more preons
bound by a knotted field.

\ve

\section{Introduction}

The possibility that the elementary particles are knots has been
suggested by many authors, going back as far as Kelvin.$^1$  Among
the different field theoretic attempts to construct classical
knots, a model related to the Skyrme soliton has been described
by Fadeev and Niemi.$^2$  There are also the familiar 
knots of a magnetic field; and since these are macroscopic
expressions of the electroweak field, it is natural to
extrapolate from macroscopic to microscopic knots of this 
same field.
One expects that the conjectured microscopic knots would be
quantized, and that they
would be observed as solitonic in virtue of both their
topological and quantum stability.  It is then natural to ask if 
the elementary particles might also be knots.  If they are, one expects
that the most elementary particles, namely the fermions, are also the
most elementary knots, namely the trefoils.  This possibility is
suggested by the fact that there are 4 
quantum trefoils and 4 classes of
elementary fermions, and is supported by a unique one-to-one 
correspondence between the topological description of the 4 quantum
trefoils and the quantum numbers of the 4 fermionic classes.  We
have first attempted to determine the minimal restrictions on a model
of the elementary particles in the context of weak interactions if the knotted soliton (quantum knot) is described only by its
symmetry algebra $SL_q(2)$ independent of its field theoretic
origin.  The use of this symmetry algebra to define the quantum
knot is similar to the use of the symmetry algebra of
the rotation group to define the quantum spin.
Before describing the symmetry algebra $SL_q(2)$ we shall describe an
oriented knot by its topological invariants and by an invariant
polynomial.

\vskip.5cm

\section{The Characterization of Oriented Knots}

Three-dimensional knots are described in terms of their projections onto a two-dimensional plane where they appear as two-dimensional curves with 4-valent vertices.  At each vertex
(crossing) there is an overline and an underline.  We shall be
interested here in oriented knots.  The crossing sign of the
vertex is +1 or -1 depending on whether the orientation of the
overline is carried into the orientation of the underline by a 
counterclockwise or clockwise rotation, respectively.  The sum of
all crossing signs is termed the writhe, $w$, a topological
invariant.  There is a second topological invariant, the 
rotation, $r$, the number of rotations of the tangent in going
once around the knot.

Let $K$ and $K^\prime$ be oriented knot diagrams with the same
writhe and rotation
\begin{eqnarray*}
w(K) &=& w(K^\prime) \\
r(K) &=& r(K^\prime)
\end{eqnarray*}
Then $K$ is topologically equivalent (regularly isotopic) to
$K^\prime$.

We may label an oriented knot by the number of crossings $(N)$,
its writhe $(w)$, and rotation $(r)$.  The writhe and rotation 
are integers of opposite parity.

\vskip.5cm

\section{The Kauffman Algorithm for Associating a Polynomial with
a Knot$^3$}

Denote the Kauffman polynomial associated with a knot, $K$, having
$n$ crossings, by $\langle K\rangle_n$.  Let us represent
$\langle K\rangle_n$ by the bracket

\be
\langle K\rangle_n \sim
\left\langle\matrix{\ldots \cr {\large\times}} \right\rangle
\ee
The interior of this bracket is intended to represent the
projected knot when only one of the $n$ crossings
is explicitly shown.  Let us also introduce the polynomials
$\langle K_\pm\rangle_{n-1}$, associated with slightly altered
diagrams in which the crossing lines are reconnected, as follows:
\be
\langle K_-\rangle_{n-1} \sim 
\left\langle\matrix{\ldots \cr \asymp}\right\rangle \quad
{\rm and} \quad \langle K_+\rangle _{n-1} \sim
\left\langle\matrix{\ldots \cr \left. \right) \! \left(
\right.}\right\rangle
\ee
Then one may define a Laurent polynomial in the parameter $q$
by the following recursive rules:
\begin{eqnarray*}
\langle K\rangle_n &=& i\left[q^{-1/2}\langle K_-\rangle_{n-1} -
q^{1/2}\langle K_+\rangle_{n-1}\right] \hspace{2.2in} (I) \\
\langle OK\rangle &=& (q+q^{-1})\langle K\rangle
\hspace{3.4in} (II) \\
\langle O\rangle &=& q+q^{-1} \hspace{3.7in} (III)
\end{eqnarray*}
Every time Rule I is applied, one crossing is eliminated and
the number of diagrams is doubled.  After $n$ applications of
Rule I, we find 2$^n$ diagrams, each with one or more internal
loops.  When these loops, indicated by O, are removed by Rules
II and III we are left with a Laurent polynomial in $q$.  Then
$\langle K\rangle_n$ is the Kauffman polynomial associated with the knot with $n$ crossings denoted by $K_n$.

The Kauffman rules may be written entirely in terms of 
the Pauli matrices $\sigma_\pm$ and the
following matrix:
\be
\epsilon_q = \left(
\begin{array}{cc}
0 & q_1^{1/2} \\ -q^{1/2} & 0
\end{array} \right)  \qquad q_1 = q^{-1}
\ee
These rules then read as follows:
\begin{eqnarray*}
\langle K\rangle_n &=& {\rm Tr}~\epsilon_q\left[\sigma_-
\langle K_-\rangle_{n-1} + \sigma_+\langle K_+\rangle_{n-1}\right]
\hspace{2.06in} (I)^\prime \\
\langle OK\rangle &=& {\rm Tr}~\epsilon_q^t\epsilon_q
\langle K\rangle \hspace{3.5in}  (II)^\prime \\
\langle O\rangle &=& {\rm Tr}~\epsilon_q^t\epsilon_q
\hspace{3.7in} (III)^\prime
\end{eqnarray*}
where
\[
\sigma_\pm = \frac{1}{2} (\sigma_1\pm i~\sigma_2)
\]
and the $\vec\sigma$ are the Pauli matrices.

One may
obtain an invariant of ambient isotopy by 
forming$^{3,4}$
\be
f_K(A) = (-A^3)^{-w(K)} \langle K\rangle
\ee
where $w(K)$ is the writhe of $K$ and
\be
A = i~{\rm Tr}~\epsilon_q\sigma_-
\ee
The Jones polynomial is 
\be
V_K(t) = f_K(t^{-1/4})
\ee

The Kauffman and Jones polynomials are topological invariants.
They are invariants of regular and ambient isotopy respectively.

\vskip.5cm

\section{The Knot Algebra$^{4,5,6}$}

The description of the knot by $(I)^\prime, (II)^\prime,
(III)^\prime$ is invariant under the transformations
\begin{subequations}
\begin{eqnarray}
\epsilon_q^\prime &=& T\epsilon_q T^t = T^t\epsilon_qT \\
\vec\sigma^\prime &=& \vec\sigma
\end{eqnarray}
\end{subequations}
where
\be
T = \left(
\begin{array}{cc}
a & b \\ c & d
\end{array} \right)
\ee
if the matrix elements of $T$ satisfy the following algebra:
\[
\left. 
\begin{array}{l}
ab=qba \\ ac=qca 
\end{array} \right.  \qquad \left.
\begin{array}{l}
bd=qdb \\ cd=qdc 
\end{array} \right. \qquad \left.
\begin{array}{l}
ad-qbc=1 \\ da-q_1cb=1
\end{array} \right. \qquad
bc=cb    \hskip1.0in (A)
\]
Then
\be
T\epsilon_q T^t = T^t\epsilon_qT = \epsilon_q
\ee
and by (4.1a)
\be
\epsilon_q^\prime = \epsilon_q
\ee

Therefore the Kauffman algorithm as expressed in terms of $\epsilon_q$
is invariant under (4.1).
We shall refer to $(A)$ as the knot algebra.  The matrix $T$, as
defined by (4.2) and $(A)$, is a 2-dimensional representation
of $SL_q(2)$.

We shall also introduce the unitary algebra $SU_q(2)$ obtained
by setting
\be
\begin{array}{rcl}
d &=& \bar a \\
c &=& -q_1\bar b
\end{array}
\ee
Then $(A)$ reduces to
\[
\left.
\begin{array}{l}
ab=qba \\ a\bar b = q\bar ba 
\end{array} \right. \quad  \left.
\begin{array}{l}
a\bar a+b\bar b = 1 \\ \bar aa+q^2_1\bar bb=1
\end{array} \right.  \quad
b\bar b=\bar bb   \hskip1.0in (A)^\prime
\]

The Kauffmann and Jones knot poynomials are left invariant by
(4.1) 
subject to either $(A)$ or $(A)^\prime$, the algebras defining
the two-dimensional representation of $SL_q(2)$ and $SU_q(2)$.  
For the physical applications we need the higher dimensional
representations of $SL_q(2)$ and $SU_q(2)$.

\vskip.5cm

\section{Higher Dimensional Representations of $SL_q(2)$ and
$SU_q(2)$}

To compute
the higher dimensional representations one needs the 
$q$-binomial theorem.$^7$  This may be written in either of the
following two ways:
\begin{subequations}
\be
(A+B)^n = \sum 
\left\langle\matrix{n \cr s \cr}\right\rangle_q
B^s A^{n-s} 
\ee
or as
\be
(A+B)^n = \sum
\left\langle\matrix{ n \cr s\cr} \right\rangle_{q_1} A^s B^{n-s}
\ee
\end{subequations}
where
\be
AB=qBA \quad {\rm and} \quad q_1=q^{-1}
\ee
Here
\be
\left\langle\matrix{n \cr s \cr} \right\rangle_q =
\frac{\langle n\rangle_q!}{\langle n-s\rangle_q!
\langle s\rangle_q!} \quad {\rm with} \quad
\langle n\rangle_q = \frac{q^n-1}{q-1}
\ee
We shall use this theorem to compute the transformations on the
following class of monomials:
\begin{subequations}
\be
\psi^j_m = N^j_m x_1^{n_+}x_2^{n_-} \quad
-j\leq m \leq j
\ee
where
\begin{eqnarray}
& &\left[x_1,x_2\right] = 0 \\
& &n_\pm = j\pm m \\
& &N_m^j = \frac{1}{[\langle n_+\rangle_{q_1}!\langle 
n_-\rangle_{q_1}!
]^{1/2}}
\end{eqnarray}
\end{subequations}
when $\left(
\begin{array}{l}
x_1 \cr x_2
\end{array} \right)$
is transformed according to the 2-dimensional representations
of $SL_q(2)$ as follows:
\begin{eqnarray}
x_1^\prime &=& ax_1 + bx_2 \\
x_2^\prime &=& cx_1+dx_2
\end{eqnarray}

Here $T=\left(
\begin{array}{cc}
a & b \\ c & d
\end{array} \right)$
is the 2-dimensional representation of $SL_q(2)$ introduced in (4.2).  
Then
\be
\psi_m^{j^\prime} = N_m^j(ax_1+bx_2)^{n_+}
(cx_1+dx_2)^{n_-}
\ee
We assume that $(a,b,c,d)$ commute with $(x_1,x_2)$ so that
\begin{eqnarray}
(ax_1)(bx_2) &=& q(bx_2)(ax_1) \\
(cx_1)(dx_2) &=& q(dx_2)(cx_1)
\end{eqnarray}
By the $q$-binomial theorem
\begin{eqnarray}
\psi^{j^\prime}_m &=& N_m^j \sum^{n_+}_s
\left\langle\matrix{n_+ \cr s\cr}\right\rangle_{q_1}
(ax_1)^s (bx_2)^{n_+-s} \sum^{n_-}_t
\left\langle\matrix{n_- \cr t\cr}\right\rangle_{q_1}(cx_1)^t
(dx_2)^{n_--t} \\
&=& N^j_m \sum_{s,t}
\left\langle\matrix{n_+ \cr s\cr}\right\rangle_{q_1}
\left\langle\matrix{n_- \cr t\cr}\right\rangle_{q_1}
x_1^{s+t}x_2^{n_++n_--s-t}a^sb^{n_+-s}c^td^{n_--t} \nonumber \\
&=& N_m^j \sum_{s,t}
\left\langle\matrix{n_+ \cr s\cr}\right\rangle_{q_1}
\left\langle\matrix{n_- \cr t\cr}\right\rangle_{q_1}
a^sb^{n_+-s}c^td^{n_--t}x_1^{n_+^\prime} x_2^{n_-^\prime}
\end{eqnarray}
where
\begin{eqnarray}
n_+^\prime &=& s+t \\
n_-^\prime &=& n_++n_--s-t
\end{eqnarray}
and by (5.4c)
\be
n_+^\prime + n_-^\prime = n_++n_- = 2j
\ee
Set
\be
n_\pm^\prime = j\pm m^\prime
\ee
We may rewrite (5.11) as
\begin{eqnarray}
\psi_m^{j^\prime} &=& \sum_{s,t} \left(\frac{N_m^j}{N^j_{m^\prime}}\right) 
\left\langle\matrix{n_+ \cr s\cr}\right\rangle_{q_1}
\left\langle\matrix{n_- \cr t\cr}\right\rangle_{q_1}
\delta(s+t,n_+^\prime)a^sb^{n_+-s}c^td^{n_--t}
(N^j_{m^\prime}x_1^{n_+^\prime}x_2^{n_-^\prime}) \\
&=&\sum_{m^\prime} D^j_{mm^\prime} \psi^j_{m^\prime}
\end{eqnarray}
where
\be
D^j_{mm^\prime} = \frac{N_m^j}{N_{m^\prime}^j}\sum_{s,t}
\left\langle\matrix{n_+ \cr s\cr}\right\rangle_{q_1}
\left\langle\matrix{n_- \cr t\cr}\right\rangle_{q_1}
\delta(s+t,n_+^\prime)a^sb^{n_+-s}c^td^{n_--t}
\ee
or
\be
D^j_{mm^\prime} = \left(\frac{\langle n_+^\prime\rangle_1!
\langle n_-^\prime\rangle_1!}{\langle n_+\rangle_1!
\langle n_-\rangle_1!}\right)^{1/2} 
\sum_{\scriptstyle 0\leq s\leq n_+
\atop\scriptstyle 0\leq t\leq n_-}
\left\langle\matrix{n_+ \cr s\cr}\right\rangle_1
\left\langle\matrix{n_- \cr t\cr}\right\rangle_1
\delta(s+t,n_+^\prime)a^sb^{n_+-s}c^td^{n_--t}
\ee
where we write $\langle~\rangle_1$ for $\langle~\rangle_{q_1}$.
The corresponding representations of $SU_q(2)$ are obtained by
setting
\begin{subequations}
\begin{eqnarray}
d &=& \bar a \\
c &=& -q_1\bar b
\end{eqnarray}
\end{subequations}
Then
\be
D^j_{mm^\prime} = \left(\frac{\langle n_+^\prime\rangle_1!
\langle n_-^\prime\rangle_1!}
{\langle n_+\rangle_1!\langle n_-\rangle_1!}\right)^{1/2}
\sum_{\scriptstyle 0\leq s\leq n_+\atop\scriptstyle 0\leq t\leq n_-}
\left\langle\matrix{n_+ \cr s\cr}\right\rangle_1
\left\langle\matrix{n_-\cr t\cr}\right\rangle_1
\delta(s+t,n_+^\prime)(-q_1)^t
a^sb^{n_+-s}\bar b^t \bar a^{n_--t}
\ee
For both $SL_q(2)$ and $SU_q(2)$ we have
\be
\psi_m^j(x_1^\prime,x_2^\prime) = \sum D^j_{mm^\prime}
\psi^j_{m^\prime}(x_1,x_2)
\ee

In obtaining these representations of $SL_q(2)$ and $SU_q(2)$ that
operate on the monomial basis (5.4a) we are following a well known
procedure for obtaining representations of $SU(2)$.$^8$

\vskip.5cm

\section{The Gauge Group of the $SL_q(2)$ and $SU_q(2)$ \\
Algebras}

By (5.21) the $2j+1$-dimensional representations of $SL_q(2)$ have the
following form
\be
D^j_{mm^\prime} = \sum_{\stackrel{0\leq s\leq n_+}{0\leq t\leq n_-}}
{\cal{A}}^j_{mm^\prime}(q,s,t)\delta(s+t,n_+^\prime)
a^sb^{n_+-s}c^t
d^{n_--t}
\ee
where $(a,b,c,d)$ satisfy the knot algebra $(A)$ defined
in Section 4.  Here
\begin{eqnarray}
n_\pm &=& j\pm m  \\ n_\pm^\prime 
&=& j\pm m^\prime
\end{eqnarray}
$D^j_{mm^\prime}$ is defined only up to the following 
gauge transformation on $(a,b,c,d)$ that leaves the algebra $(A)$
invariant:
\begin{center}
\begin{tabular}{ll}
$a^\prime = e^{i\varphi_a}a$  \qquad & $b^\prime = e^{i\varphi_b}b$ \\
$d^\prime = e^{-i\varphi_a}d$ \qquad & $c^\prime = e^{-i\varphi_b}c$
\hspace{2.75in} $(G)$
\end{tabular}
\end{center}
We shall also refer to $(G)$ as $U_a(1)\times U_b(1)$.  Under
the gauge transformation $(G)$, every term in
$D^j_{mm^\prime}$ transforms like
\be
(a^{n_a}b^{n_b}c^{n_c}d^{n_d})^\prime =
e^{i\varphi_a(n_a-n_d)}e^{i\varphi_b(n_b-n_c)}
(a^{n_a}b^{n_b}c^{n_c}d^{n_d})
\ee
But by the $\delta$-function in (6.1)
\be
\begin{array}{rcl}
n_a-n_d &=& s+(t-n_-)=n_+^\prime-n_- = m^\prime+m \\
n_b-n_c &=& (n_+-s)-t = n_+-n_+^\prime = m-m^\prime
\end{array}
\ee
By (6.4) and (6.5) every term of $D^j_{mm^\prime}$
transforms the same way and therefore the 
$D^j_{mm^\prime}$
transforms under $G$ as follows:
\begin{subequations}
\be
D^{j~\prime}_{mm^\prime} = e^{i(m+m^\prime)\varphi_a}
e^{i(m-m^\prime)\varphi_b}D^j_{mm^\prime} 
\ee
or
\be
D^{j~\prime}_{mm^\prime} = e^{i(\varphi_a+\varphi_b)m}
e^{i(\varphi_a-\varphi_b)m^\prime}D^j_{mm^\prime}
\ee
\end{subequations}
We denote the irreducible representations of 
$SU_q(2)$ by $D^j_{mm^\prime}(a,\bar a,b,\bar b)$.

The gauge transformations on $SU_q(2)$, namely
\be
\begin{array}{rcl}
a^\prime &=& e^{i\varphi_a}a  \\
b^\prime &=& e^{i\varphi_b}b
\end{array}
\ee
induce the same transformations (6.6) on the $D^j_{mm^\prime}
(a,\bar a,b,\bar b)$.

\vskip.5cm

\section{Representation of an Oriented Knot}

The oriented knot has three coordinates, namely $(N,w,r)$ the
number of crossings $N$, the writhe $w$, and the rotation $r$.
We may make a coordinate transformation to $(j,m,m^\prime)$,
the indices that label the irreducible representations
$D^j_{mm^\prime}$ of $SL_q(2)$ by setting
\be
\begin{array}{rcl}
j &=& N/2 \\
m &=& w/2 \\
m^\prime &=& (r+1)/2
\end{array}
\ee
This linear transformation allows half-integer representations
and respects the knot constraint requiring $w$ and $r$ to be of
opposite parity.  In this new coordinate system one may label
the knot $(N,w,r)$ by $D^{N/2}_{\frac{w}{2}\frac{r+1}{2}}
(a,b,c,d)$.  One thereby associates with the $(N,w,r)$ knot a 
multinomial in the elements of the algebra of the form
\be
D^j_{mm^\prime}(abcd) = \sum {\cal{A}}^j_{mm^\prime}
(q,s,t)\delta(s+t,n_+^\prime) a^sb^{n_+-s}c^td^{n_--t}
\ee
where explicit forms of ${\cal{A}}^j_{mm}$ are given in (5.19)
and (5.21). 

Like the Kauffman and Jones polynomials these forms
are based on the algebra of the classical knot.  They are
operator expressions that may be evaluated on the state space of
the algebra.

Let us next compute a basis in this space.

Since $b$ and $c$ commute, they have common eigenstates.  Let
$|0\rangle$ be designated as a ground state and let
\begin{eqnarray}
b|0\rangle &=& \beta|0\rangle \\
c|0\rangle &=& \gamma|0\rangle \\
bc|0\rangle &=& \beta\gamma|0\rangle
\end{eqnarray}

We may assume that $b$ and $c$ are Hermitian:
\begin{eqnarray}
b &=& \bar b \\
c & = \bar c
\end{eqnarray}
Then the eigenvalues $\beta,\gamma$ are real and the eigenfunctions are orthogonal.

From the algebra we see that
\be
bc|n\rangle = E_n|n\rangle
\ee
where
\be
|n\rangle \sim d^n|0\rangle
\ee
and
\be 
E_n = q^{2n}\beta\gamma
\ee
This eigenvalue spectrum resembles that of a harmonic oscillator
but the levels are arranged in geometrical rather than
arithmetical progression.  We shall refer to this spectrum as
the $q$-oscillator spectrum.

Here $d$ and $a$ are raising and lowering operators respectively.
\begin{eqnarray}
d|n\rangle &=& \lambda_n|n+1\rangle \\
a|n\rangle &=& \mu_n|n-1\rangle
\end{eqnarray}
If there is a highest state $M$, $\lambda_M=0$; if there is a
lowest state $M^\prime$, $\mu_{M^\prime} = 0$.
We also have
\be
\begin{array}{rcl}
ad|n\rangle &=& a\lambda_n|n+1\rangle \\
&=& \lambda_n\mu_{n+1}|n\rangle 
\end{array}
\ee
\be
\begin{array}{rcl}
da|n\rangle &=& d\mu_n|n-1\rangle \\
&=& \mu_n\lambda_{n-1}|n\rangle
\end{array}
\ee
From the algebra $(A)$, (7.13) and (7.14) become
\begin{eqnarray}
(1+qbc)|n\rangle &=& \lambda_n\mu_{n+1}|n\rangle \\
(1+q_1bc)|n\rangle &=& \mu_n\lambda_{n-1}|n\rangle
\end{eqnarray}
If there is both a highest state $M$, and a lowest state 
$M^\prime$, then
\be
\lambda_M = \mu_{M^\prime} = 0 \qquad M^\prime < M
\ee
and by (7.15) and (7.16)
\begin{eqnarray}
(1+qbc)|M\rangle &=& 0 \\
(1+q_1bc)|M^\prime\rangle &=& 0
\end{eqnarray}
Then by (7.8) and (7.10)
\be
q^{2M+1}\beta\gamma = q^{2M^\prime-1}\beta\gamma
\ee
or
\be
(q^2)^{M-M^\prime+1}=1
\ee
We assume that $q$ is real, so that
\be
M^\prime = M+1
\ee

Since (7.17) and (7.22) are contradictory, there may be either
a highest or a lowest state but not both.  The same discussion
may be given for the $SU_q(2)$ algebra.

In the next section we shall assume
that the individual states of excitation of the quantum
knot are to be represented by $D^j_{mm^\prime}|n\rangle$.  Since the
empirical evidence appears to restrict the number of states, there
must be an externally required physical boundary condition to
cut off the otherwise infinite spectrum that is formally
required by (7.10).

\vskip.5cm

\section{The Quantum Knot$^{9,10,11,12}$}

Since the writhe and rotation of a classical knot
are regular topological invariants,
they do not depend on the size or shape of the knot; i.e., they are 
conformal invariants that hold for microscopic knots as well.  It
follows that $w$ and $r$ are integrals of the motion for microscopic
classical knots with spectra determined by the topology of the knot.

We shall now introduce the quantum knot by interpreting
$D^j_{mm^\prime}(a,b,c,d)$ as the kinematical description of a quantum
state, where
\be
(j,m,m^\prime) = \frac{1}{2}(N,w,\pm r+1)
\ee
and $(N,w,r)$ describes a classical knot.  Since the spectra of
$(j,m,m^\prime)$ are restricted by $SL_q(2)$, and the spectra of
$(N,w,r)$ are restricted by knot topology, the states of the
quantized knot are thus jointly restricted by both $SL_q(2)$ and the
knot topology.  The equations (8.1) establish a correspondence
between a quantized knot described by $D^{N/2}_{\frac{w}{2}
\frac{\pm r+1}{2}}$ and a classical knot described by $(N,w,r)$, but
the correspondence is not one-to-one.  There is a one-to-one
correspondence between the quantum trefoil and the 2d-projection.

For the trefoil configuration
there are four choices of $(w,r)$, namely (3,2),
(-3,2), (3,-2), (-3,-2).  Regarded as classical knots, only two of
these trefoils are topologically distinct; but we shall consider all
four choices of $D^{3/2}_{\frac{w}{2}\frac{r+1}{2}}$ as distinct
quantum states, since the ``topological degeneracy" is lifted by
turning on the hypercharge, as we shall see.  In the following when
$(w,r)$ refers to the quantum knot, both $w$
and $r$ may have either sign.

One may similarly define the eigenstates of the spherical top
as irreducible representations of $O(3)$ by 
$D^j_{mm^\prime}(\alpha,\beta,\gamma)$ where the indices
$(j,m,m^\prime)$ refer to the angular momentum of the top and the
arguments $(\alpha,\beta,\gamma)$ to its orientation.  It is
also possible to define the eigenstates of the hydrogen atom as
irreducible representations of $O(3)$, expressed as
$D^j_{mm^\prime}(a_1,a_2,a_3)$ where in this case
$(a_1,a_2,a_3)$ are three coordinates on the group space of
$O(3)$, and where $(2j+1,m,m^\prime)$ are respectively the
principal quantum number, the $z$-component of the angular
momentum, and the $z$-component of the Runge-Lenz vector.$^{9}$  
Here
the quantum knot is similarly described, but it is defined on the
$SL_q(2)$ algebra, a discrete space rather than a three-dimensional continuum.

If the knot oscillates
like a quantum mechanical harmonic oscillator, the Hamiltonian is
of the following form:
\begin{subequations}
\be
H = (a\bar a + \bar aa) \frac{\hbar\omega}{2}
\ee
where $\bar a$ and $a$ are raising and lowering operators and
\be
[a,\bar a] = 1
\ee
\end{subequations}

Since the raising and lowering operators of the $SL_q(2)$ algebra
that correspond to $\bar a$ and $a$ of the harmonic oscillator
are $d$ and $a$, the knot analogue of (8.2) is
\begin{subequations}
\be
H = (ad + da)\frac{\hbar\omega}{2}
\ee
but
\be
[a,d] = (q-q_1) bc
\ee
\end{subequations}
and
\be
\frac{1}{2} (ad+da) = 1 + \frac{1}{2} (q+q_1)bc
\ee
We may generalize the $SL_q(2)$ analog of the harmonic oscillator  
by replacing (8.3) by a
more general function of $ad+da$, namely:
\be
H = H(ad+da) \frac{\hbar\omega}{2}
\ee 
or by (8.4) with a different $H$
\be
H = H(bc) \frac{\hbar\omega}{2}
\ee
or with a still different $H$ 
\be
H = H(b,c) \frac{\hbar\omega}{2}
\ee

Let the Hamiltonian of a quantum knot be 
$H(b,c)$.  
Let us consider
the states of this knot defined by 
$D^j_{mm^\prime}|n\rangle$. We may then compute
\begin{eqnarray}
H(b,c)D^j_{mm^\prime}|n\rangle &=& H(b,c)\left[
\sum_{s,t}{\cal{A}}^j_{mm^\prime}\delta(s+t,n_+^\prime)a^s
b^{n_+-s}c^td^{n_--t}\right]|n\rangle \\
&=& D^j_{mm^\prime}H(q_1^{n_a-n_d}b,q_1^{n_a-n_d}c)|n\rangle
\end{eqnarray}
where $n_a$ and $n_d$ are the exponents of $a$ and $d$
respectively, and $n_a-n_d$ is by (6.5) the same for every
term of $D^j_{mm^\prime}$.  Then one has by (7.3), (7.4), and (7.10)
\be
\begin{array}{rcl}
H(b,c)D^j_{mm^\prime}|n\rangle &=& D^j_{mm^\prime}
H(q_1^{n_a-n_d}q^n\beta,q_1^{n_a-n_d}q^n\gamma|n\rangle \\
&=& E^j_{mm^\prime}(n){\cal{D}}^j_{mm^\prime}|n\rangle
\end{array}
\ee
where the eigenvalues of $H$ are
\be
E^j_{mm^\prime}(n) = H(\lambda\beta,\lambda\gamma)
\ee
and
\be
\lambda = q^{n-(m+m^\prime)} 
\ee
by (6.5).  The eigenstates of $H$ are the 
$D^j_{mm^\prime}|n\rangle$ and
the indices on $D^j_{mm^\prime}$ are the 
eigenvalues of the integrals of motion.

The operators that represent the integrals of the motion may be
expressed in terms of an elementary operator $\omega_x$ that may
be defined by its action on every term of 
$D^j_{mm^\prime}$ as follows:
\be
\omega_x(\ldots x^{n_x}\ldots) = n_x(\ldots x^{n_x}\ldots) \qquad
~~~~x = (a,b,c,d)
\ee
i.e., $\omega_x$ acts like $x\frac{\partial}{\partial x}$, a
dilatation operator.

Then define
\begin{eqnarray}
{\cal{N}} &=& (\omega_a+\omega_b + \omega_c +
\omega_d) \\
{\cal{W}} &=&  (\omega_a - \omega_d + \omega_b -
\omega_c) \\
{\cal{R}} &=&  (\omega_a-\omega_d - \omega_b +
\omega_c)
\end{eqnarray}
When ${\cal{N}},{\cal{W}}$, and ${\cal{R}}$ act on
$D^j_{mm^\prime}$ one finds by (6.1)
\begin{eqnarray}
{\cal{N}}~D^j_{mm^\prime} &=& 
2j~D^j_{mm^\prime} \\
{\cal{W}}~D^j_{mm^\prime} &=& 
2m~D^j_{mm^\prime} \\
{\cal{R}}~D^j_{mm^\prime} &=& 
2m^\prime~D^j_{mm^\prime}
\end{eqnarray}
We shall describe a state function of the quantum knot by 
\be
\psi^N_{wr} = D^{N/2}_{\frac{w}{2}\frac{r+1}{2}}
|n\rangle
\ee
following (7.1) where $(N,w,|r|)$ are the number of crossings, writhe,
and rotation of the corresponding classical knot.
Then by Eqs. (8.17)-(8.19) we have
\begin{eqnarray}
{\cal{N}}~\psi^N_{wr} &=& N~\psi^N_{wr} \\
{\cal{W}}~\psi^N_{wr} &=& w~\psi^N_{wr} \\
{\cal{R}}~\psi^N_{wr} &=& (r+1)~\psi^N_{wr}
\end{eqnarray}
where the spectra of $({\cal{N}},{\cal{W}},{\cal{R}})$ are restricted
by the topology of the knot.
In addition we have
\be
H(b,c)\psi^N_{wr} = E^N_{wr} \psi^N_{wr}
\ee
where by (8.12) and (7.1)
\begin{subequations}
\begin{eqnarray}
E^N_{wr} &=& H(\lambda\beta,\lambda\gamma) \qquad
{\rm and} \\
\lambda &=& q^{n-\frac{1}{2}(w+r+1)}
\end{eqnarray}
\end{subequations}

\section{The Quantum Knot and the Standard Theory$^{10,11,12}$}

One may now attempt to relate 
$D^{N/2}_{\frac{w}{2}\frac{r+1}{2}}(q|abcd)$ to the internal
state of an elementary particle, which we shall
assume to be a boson if $N$ is even
and a fermion if $N$ is odd.  Since the lowest value of $N$
is 3, corresponding to a trefoil, we shall try to identify
the four quantum trefoils with the four classes of elementary
fermions, namely
\[
\begin{array}{cccc}
(1) & \nu_e & \nu_\mu & \nu_\tau \\
(2) & e & \mu & \tau \\
(3) & d & s & b \\
(4) & u & c & t
\end{array}
\]
We shall assume that each quantum trefoil has 3 states of
excitation, e.g., the 3 states of the leptonic trefoil represent
$(e,\mu,\tau)$.  We shall now represent all four of the elementary fermionic trefoils by $D^{3/2}_{\frac{w}{2}
\frac{r+1}{2}}$ and the three states of each trefoil by
$D^{3/2}_{\frac{w}{2}\frac{r+1}{2}}|n\rangle$, $n=0,1,2$.

In order to identify the 4 quantum trefoils with the 4
families of fermions, it is necessary to establish a unique
correspondence between the 4 choices of writhe and rotation 
that label the quantum trefoils and the 4 choices of charge and
hypercharge that distinguish the 4 families of fermions.  For
this purpose we introduce two ``knot charges" $Q_a$ and $Q_b$
by rewriting (6.6) as follows:

Let
\begin{eqnarray}
Q_a &\equiv& -k(m+m^\prime) = -k \frac{w+r+1}{2} \\
Q_b &\equiv& -k(m-m^\prime) = -k \frac{w-r-1}{2}
\end{eqnarray}
where $k$ is an undetermined constant with the dimensions of an
electric charge.
The classical
$Q_a$ and $Q_b$ are conserved since $w$ and $r$ are conserved.
By (6.6)
the gauge transformations $(G)$ on the algebra $(A)$ 
induce the following gauge transformations on the kinematical states
\be
D^{j^\prime}_{mm^\prime} = U_aU_bD^j_{mm^\prime}
\ee
where
\begin{eqnarray}
U_a &=& e^{-ik^{-1}Q_a\varphi_a} \\
U_b &=& e^{-ik^{-1}Q_b\varphi_b}
\end{eqnarray}
Then $U_a$ and $U_b$ are two independent gauge transformations
on the knot states, and may be compared with the two independent
gauge transformations defining charge and hypercharge.  To
examine this correspondence we compare the knot charges $Q_a$
of the 4 quantum trefoils with the electric charges $Q_f$ of the four
families in Table 9.1.

\vskip.3cm
\ce{\bf Table 9.1}
\vskip.3cm

\[
\begin{array}{ccc|cc} \hline
{\rm Trefoils}~(w,r) \quad & D^{N/2}_{\frac{w}{2}\frac{r+1}{2}}
\quad & Q_a \quad & {\rm Fermion~Class} \quad & Q_f \\ \hline
(-3,2) \quad & D^{3/2}_{-\frac{3}{2}\frac{3}{2}} \quad & 0
\quad & (\nu_e\nu_\mu\nu_\tau) \quad & 0 \\
(3,2) \quad & D^{3/2}_{\frac{3}{2}\frac{3}{2}} \quad & -3k
\quad & (e^-,\mu^-,\tau^-) \quad & -e \\
(3,-2) \quad & D^{3/2}_{\frac{3}{2}-\frac{1}{2}} \quad & -k
\quad & (d,s,b) \quad & -\frac{1}{3}e \\
(-3,-2) \quad & D^{3/2}_{-\frac{3}{2}-\frac{1}{2}} \quad & 2k
\quad & (u,c,t) \quad & \frac{2}{3}e \\ \hline
\end{array}
\]
In Table (9.1) and by Eq. (9.1)
\be
\begin{array}{rcl}
Q_a &=& -k(m+m^\prime) = -\frac{k}{2} (w+r+1) \\
Q_f &=& \mbox{electric charge of fermion class}
\end{array}
\ee
\underline{There is a unique mapping and 
single value of $k$} that permits 
one to match the trefoil knots with the fermion classes by
satisfying
\be
Q_a(w,r) = Q_f
\ee
where $k$ appears as the quantum of charge:
\be
k = \frac{e}{3}
\ee
Then
\be
Q_a = -\frac{e}{6} (w+r+1)
\ee
may be considered the electric charge of the quantum trefoil.

The above mapping is unique in the sense that any other
correspondence between the trefoils and the fermion classes
would destroy the proportionality between $Q_a$ and $Q_f$ and would
therefore require more than a single value of $k$.

Since $Q_a \sim m+m^\prime=n_a-n_d$ by (9.1) and
(6.5), note that the vanishing
of $Q_a$ implies
\be
n_a = n_d
\ee
and therefore that $a$ and $d$ may be eliminated from
every term of $D^j_{mm^\prime}$ with the aid of
\be
a^nd^n = \prod^n_{s=1} (1+q^{2s-1}bc)
\ee
as follows from 
the $SL_q(2)$ algebra $(A)$.  Therefore electrically neutral
states (neutrinos and neutral bosons) lie entirely in the
$(b,c)$ subalgebra.

If the symmetry group is $SU_q(2)$ then the neutral states lie
in the $(b,\bar b)$ subalgebra.
Also $\bar D^j_{mm^\prime}$ has opposite charges from
$D^j_{mm^\prime}$ and may be identified as the state
of the antiparticle.

Given the match in Table (9.1) we may now compare all the quantum
numbers $(t,t_3,Q)$ labeling the different classes of fermions
in the standard representation with the quantum numbers
$(N,w,r)$ labeling the corresponding quantum trefoils.

\ve

\ce{\bf Table 9.2} 

\vskip.2cm

\begin{center}
\begin{tabular}{lccc||cccc} \hline
\multicolumn{4}{c}{Standard Representation} & \multicolumn{4}{c}
{Trefoil Representation} \\ \hline
& $t$ & $t_3$ & $Q$ & 
$w$ & $r$ & 
$D^{3/2}_{\frac{w}{2}\frac{r+1}{2}}$
& $Q_a$ \\ \hline
$(e\mu\tau)_{\rm L}$ & $\frac{1}{2}$ & $-\frac{1}{2}$ & $-e$ & 
$3$ & $2$ & $D^{3/2}_{\frac{3}{2}\frac{3}{2}}$ & $-e$ \\
$(\nu_e\nu_\mu\nu_\tau)_{\rm L}$ & $\frac{1}{2}$ & $\frac{1}{2}$ & $0$ &
$-3$ & $2$ & $D^{3/2}_{-\frac{3}{2}\frac{3}{2}}$ & $0$ \\
$(dsb)_{\rm L}$ & $\frac{1}{2}$ & $-\frac{1}{2}$ & $-\frac{1}{3}~e$ &
$3$ & $-2$ & $D^{3/2}_{\frac{3}{2}-\frac{1}{2}}$ & $-\frac{1}{3}~e$ \\
$(uct)_{\rm L}$ & $\frac{1}{2}$ & $\frac{1}{2}$ & $\frac{2}{3}~e$ &
$-3$ & $-2$ & $D^{3/2}_{-\frac{3}{2}-\frac{1}{2}}$ & $\frac{2}{3}~e$ \\ \hline
\end{tabular}
\end{center}

\vskip.3cm

One then reads off the following relations from Table 9.2.
\be
t = \frac{N}{6}
\ee
since $N=3$ for trefoils.  

Also $t_3$ is proportional to $w$ (not to $r$) and
\be
t_3 = -\frac{w}{6}
\ee
Since $m = \frac{w}{2}$
\be
t_3 = -\frac{m}{3}
\ee

Finally in the knot representation the electric charge is by (9.1)
and (9.8)
\be
Q_a = -\frac{e}{3}~(m+m^\prime)
\ee
But in the standard theory (point particle representation)
\be
Q = (t_3+t_0) e
\ee

Since (9.15) and (9.16) must agree, we have
\be
t_3 + t_0 = -\frac{1}{3}~(m+m^\prime)
\ee
By (9.14) and (9.17) the hypercharge is
\be
t_0 = -\frac{1}{3}~m^\prime
\ee
Therefore alternative forms of the quantum state of the fermionic
knots are
\be
D^{N/2}_{\frac{w}{2}\frac{r+1}{2}} \qquad \mbox{or} \qquad
D^{3t}_{-3t_3~-3t_0}
\ee
Therefore the invariance group of the algebra, namely $U_a(1)\times
U_b(1)$, defines the charge and hypercharge.  

The preceding relations may be summarized as follows:
\be
\addtocounter{equation}{1} 
\left\{\begin{array}{l}
t=\frac{N}{6} \\ t_3=-\frac{w}{6} \\ t_0=-\frac{r+1}{6} \\
Q_e=-\frac{e}{6}(w+r+1) \end{array} \right.  
~ \quad {\rm (9.20)} ~ \quad {\rm or} 
\qquad \left\{ \begin{array}{l}
t = \frac{j}{3} \\ t_3=-\frac{m}{3} \\ t_0 = -\frac{m^\prime}{3} \\
Q_e = -‭\frac{e}{3}(m+m^\prime) \end{array} \right.
\ee

For trefoils $N=3$ and $j=\frac{N}{2} = \frac{3}{2}$.  The factor
$\frac{1}{3}$ appears because $N=3$ and independently because
the quantum of charge is $e/3$.
The additional factor of 1/2 in the 1/6 factor appears because we are describing fermions by 1/2 integer representations.

Note also that
\be
Q_e = -\frac{e}{N} \left(\frac{w+r+1}{2}\right)\quad \mbox{or}
\quad -e\left(\frac{m+m^\prime}{2j}\right) \quad \mbox{and}
\quad D^j_{mm^\prime} = D^{3t}_{-3t_3-3t_0}
\ee
hold for all the fermionic knots.
 
We may also define the ``writhe charge" and the ``rotation charge":
\begin{eqnarray}
Q(w) &\equiv& -k\frac{w}{2} \\
Q(r) &\equiv& -k\frac{r+1}{2}
\end{eqnarray}
Then we have
\begin{eqnarray}
Q_a &=& Q(w) + Q(r) \\
Q_b &=& Q(w) - Q(r) \\
Q_e &=& Q(w) + Q(r) 
\end{eqnarray}
and
\begin{eqnarray}
Q(w) &=& et_3 \\
Q(r) &=& et_0
\end{eqnarray}
i.e. $t_3$ and $t_0$ measure the ``writhe charge" and ``rotation 
charge" respectively
and the electric charge measures the sum.

The earlier equations (6.6b) and (9.3) may be rewritten in
terms of $Q(w)$ and $Q(r)$ as
\be
D^{N/2}_{\frac{w}{2}\frac{r+1}{2}}(a^\prime b^\prime
c^\prime d^\prime) = e^{-\frac{i}{k} Q(w)\varphi_w}
e^{-\frac{i}{k}Q(r)\varphi_r} 
D^{N/2}_{\frac{w}{2}\frac{r+1}{2}}(abcd)
\ee

Finally, by (8.18), (8.19) and (9.21) we define the charge operator
\be
Q = -\frac{e}{3}\frac{{\cal{W}}+{\cal{R}}}{2}
\left(= -\frac{e}{3}(m+m\prime)\right)
\ee
and by (8.15) and (8.16)
\begin{eqnarray}
{} &=& -\frac{e}{3} (\omega_a-\omega_d) 
\qquad \mbox{with}~SL_q(2) \\
{} &=& -\frac{e}{3}(\omega_a-\omega_{\bar a}) \qquad \mbox{with}~
SU_q(2)
\end{eqnarray}
and by (9.19) and (9.31)
\be
Q~D^{3t}_{-3t_3-3t_0} = e(t_3+t_0) D^{3t}_{-3t_3 -3t_0}
\ee

\vskip.5cm

\section{The Fermion-Boson Interactions$^{11,12}$}

To discuss interactions we introduce a knot field theory by
replacing the point particles of standard theory with quantum
knots.  This is done by attaching to each normal mode a knot
state just as one introduces spin by attaching a spin state.
The knot states and therefore the corresponding fields are
represented by
operators defined only up to the gauge transformations 
(9.4) and (9.5), and we
shall require the action to be invariant under these gauge
transformations, since they are induced by the
underlying transformations
that leave the defining algebra invariant.  Therefore, by
Noether's theorem, $Q(w)$ and $Q(r)$ behave, in the
field theory as well, as conserved
charges, consistent with their identification as topological
charges.

We therefore assume that the topological charges
associated with the knot gauge group are conserved by the
emission and absorption of bosonic solitons, which also carry
topological charge, as a consequence of the following
fermion-boson interaction.
\be
\bar{\cal{F}}_3{\cal{B}}_2{\cal{F}}_1
\ee
where
\begin{eqnarray}
{\cal{F}}_1 &=& F_1(p,s,t)
D^{3/2}_{m_1m_1^\prime}(abcd)
|n_1\rangle \\
\bar{\cal{F}}_3 &=& \langle n_3|\bar
D^{3/2}_{m_3m_3^\prime}(abcd)\bar F_3(p,s,t) \\
{\cal{B}}_2 &=& B_2(p,s,t)
D^j_{m_2m_2^\prime}(abcd)
\end{eqnarray}
Here $F(p,s,t)$ and $B(p,s,t)$ are the standard fermionic and
bosonic normal modes where $p$ and $s$ refer to
momentum and spin.  Then (10.1) becomes
\be
(\bar F_3B_2F_1)\langle n_3|\bar D^{3/2}_{m_3m_3^\prime}
D^j_{m_2m_2^\prime}
D^{3/2}_{m_1m_1^\prime}|n_1\rangle
\ee
The correction to the standard matrix elements appears in the
second factor, namely
\be
\langle n_3|\bar 
D^{3/2}_{m_3m_3^\prime}D^j_{m_2m_2^\prime}
D^{3/2}_{m_1m_1^\prime}|n_1\rangle
\ee
If there are $M$ generations of fermions, then $n_1$ and $n_3$
take on values $0,\ldots, M-1$.  All present evidence appears to favor
$M=3$.  We shall assume that the only occupied states $|n\rangle$ are
$n=0,1,2$ in order of increasing mass.

We require that the basic internal interaction be invariant
under gauge transformations, $U_a(1)\times U_b(1)$, of the
underlying algebra, i.e.
\be
\left(\bar D^{3/2}_{m_3m_3^\prime}\right)^\prime
\left(D^j_{m_2m_2^\prime}\right)^\prime
\left(D^{3/2}_{m_1m_1^\prime}\right)^\prime =
\bar D^{3/2}_{m_3m_3^\prime}D^j_{m_2m_2^\prime}
D^{3/2}_{m_1m_1^\prime}
\ee
Then by (9.4) and (9.5)
\begin{eqnarray}
\exp[ik\varphi_a](-Q_a(3)+Q_a(2)+Q_a(1)) &=& 1 \\
\exp[ik\varphi_b](-Q_b(3)+Q_b(2)+Q_b(1)) &=& 1
\end{eqnarray}
Therefore both $Q_a$ and $Q_b$ are conserved:
\be
Q(1)+Q(2) = Q(3)
\ee
Then by (9.1) and (9.2)
\begin{eqnarray}
m_3 &=& m_1+m_2 \\
m_3^\prime &=& m_1^\prime + m_2^\prime
\end{eqnarray}
and the possible values of $(j,m,m^\prime)$ for the intermediate
boson are restricted by the known values of $(j,m,m^\prime)$ for
the initial and final fermions.

If the rules for connecting $m$ and $m^\prime$ to $t_3$ and
$t_0$ are extended without change from fermions to the 
intermediate boson, then the conservation of $Q_a$ and $Q_b$ by
the basic interaction implies the conservation of $t_3$ and
$t_0$ by the same interaction.  Therefore we adopt for bosonic
knots the same rules as for fermionic knots:
\be
\begin{array}{rcl}
m &=& -3t_3  \\
m^\prime &=& -3t_0  \\
j &=& 3t   
\end{array}
\ee

Applied to the vector bosons these rules imply Table 10.1.  The
first three columns of Table 3 express the fact that $\vec W$ 
is an isotriplet and
$W^0$ is an isosinglet in the standard theory.

The fourth column $D^{3t}_{-3t_3~-3t_0}$ labels the internal
states of the four vector bosons.  If $t=1,j=3$; and if
$j=\frac{N}{2}$, as we have assumed, then $N=6$ and $W$ is a
ditrefoil consistent with the pair production of fermions by
vector bosons.

\vskip.5cm

\ce{\bf Table 10.1}
\vskip.3cm

\[
\begin{array}{ccrcc} \hline
\quad & t \quad & t_3 \quad & t_0 \quad &
D^{3t}_{-3t_3~-3t_0}  \\ \hline
W^+ \quad & 1 \quad & 1 \quad & 0 \quad & D^3_{-3~0} \\
W^- \quad & 1 \quad & -1 \quad & 0 \quad & D^3_{3~0} \\
W^3 \quad & 1 \quad & 0 \quad & 0 \quad & D^3_{0~0} \\
W^0 \quad & 0 \quad & 0 \quad & 0 \quad & D^0_{0~0} \\ 
\hline
\end{array} 
\]

Note
\be
Q_e = -\frac{e}{N} (w+r+1)
\ee
for the vector bosons, corresponding to (9.22) for fermions.

Since $W^0$ is a $U(1)$ coupling in the standard theory, there is
no self-coupling, i.e. it itself carries neither electric nor
hypercharge.  The assignment of $j$ to $W^0$ is also restricted
in the internal matrix element by the $q$-Clebsch-Gordan 
rules.$^{13,14}$
If $j=0$ and we maintain the relation $j=N/2$, then $N=0$ and
$W^0$ is an unknotted clockwise loop.

The possibility of extending the conservation laws and the same
rule for associating $Q_a$ and $Q_b$ with 
$m$ and $m^\prime$ to all
solitons, as here defined, depends on the fact that $Q_a$ and $Q_b$ 
depend solely on $m$ and $m^\prime$ and are independent
of $j$.

The conservation of $Q_a$ and $Q_b$, 
or equivalently of $Q_w$ and $Q_r$, i.e. the invariance of the
action under $U_a(1)\otimes U_b(1)$ is the origin in the
knot model of the conservation of $t_3$ and $t_0$.  Electric
charge and hypercharge
in this model are characterizations of
the topology of the knotted soliton.  Since we are describing a
modified standard model, it is essential that $t_3$ and $t_0$ 
defined by $U_a(1)\times U_b(1)$ agree with the $t_3$ and $t_0$
defined by $SU(2)\times U(1)$.  This agreement is expressed in
(10.13).

\vskip.5cm

\section{Fermion and Boson State Functions$^{11}$}

Acording to (10.5) and (10.6) of the knot model the standard
matrix elements are modified by the form factors (10.6).   To
compute these form factors one needs the state functions of the
fermionic knots as well as the state functions of the bosonic
knots.  They are shown in Tables 11.1 and 11.2 computed according
to (5.19) expressed as follows:
\be
D^j_{mm^\prime} = \Delta^j_{mm^\prime} \Sigma^j_{mm^\prime}
\ee
where
\be
\Delta^j_{mm^\prime} = \left[\frac{\langle n_+^\prime\rangle_1~
\langle n^\prime_-\rangle_1}{\langle n_+\rangle_1~
\langle n_-\rangle_1}\right]^{1/2}
\ee
\be
\Sigma^j_{mm^\prime} = \sum_{\scriptstyle 0\leq s\leq n_+\atop
\scriptstyle 0\leq t\leq n_-}
\left\langle\matrix{n_+ \cr s\cr}\right\rangle_1
\left\langle\matrix{n_- \cr t\cr}\right
\rangle_1\delta(s+t,n_+^\prime)
a^sb^{n_+-s}c^td^{n_--t}
\ee
Here
\[
\left\langle\matrix{n \cr s\cr}\right\rangle_1 =
\left\langle\matrix{n\cr s\cr}\right\rangle_{q_1}
\]
where $\left\langle\matrix{n\cr s\cr}\right\rangle_q$ is given
by (5.3).  In the following tables and applications we shall pass
from $SL_q(2)$ to $SU_q(2)$ by setting
\begin{eqnarray*}
d &=& \bar a \\ c &=& -q_1\bar b
\end{eqnarray*}
in (11.3).  We record these form factors for a few
of the elementary processes.

\begin{center}
{\bf Table 11.1}
\end{center}
\no {\it Fermions:}
\[
\begin{array}{rcccc}
& \underline{(e\mu\tau)} & \underline{(dsb)} & 
\underline{(uct)} & \underline{\nu_e\nu_\mu\nu_\tau)}
\\
(w,r):~~ & (3,2) ~& (3,-2) ~& (-3,-2) ~& (-3,2) \\
D^{N/2}_{\frac{w}{2}\frac{r+1}{2}}:~~ & D^{3/2}_{\frac{3}{2}
\frac{3}{2}} ~& D^{3/2}_{\frac{3}{2}-\frac{1}{2}} ~&
D^{3/2}_{-\frac{3}{2}-\frac{1}{2}} ~&
D^{3/2}_{-\frac{3}{2}\frac{3}{2}} \\
D^{N/2}_{\frac{w}{2}\frac{r+1}{2}}:~~ & a^3 ~&
\Delta^{3/2}_{-\frac{3}{2}-\frac{1}{2}}
\left\langle\matrix{3\cr 1\cr}\right\rangle_1ab^2 ~& 
-\Delta^{3/2}_{-\frac{3}{2}\frac{1}{2}}
\left\langle\matrix{3\cr 1\cr}\right\rangle_1
q_1\bar b\bar a^2 ~& -q_1^3\bar b^3
\end{array}
\]

\vskip.5cm

\begin{center}
{\bf Table 11.2}
\end{center}
\no {\it Bosons:}
\[
\begin{array}{rcccc}
& \underline{W^-} & \underline{W^+} & \underline{W^3} &
\underline{W^0} \\
D^{3t}_{-3t_3-3t_0}:~~& D^3_{30} ~& D^3_{-30} ~& D^3_{00} ~&
D^0_{00} \\
D^{3t}_{-3t_0-3t_0}:~~& \Delta^3_{30}\left\langle\matrix{6\cr
3}\right\rangle_1a^3b^3 ~& -\Delta^3_{-30}
\left\langle\matrix{6\cr 3}\right\rangle_1q_1^3\bar b^3
\bar a^3 ~& f_3(\bar bb) ~& f_0(\bar bb)
\end{array}
\]
where
\[
\Delta^3_{-30} = \Delta^3_{30} = \frac{\langle 3\rangle_1}
{\langle 6\rangle_1^{1/2}} \quad {\rm and} \quad
\left\langle\matrix{6\cr 3}\right\rangle_1 = 
\frac{\langle 6\rangle_1!}{\langle 3\rangle_1!~
\langle 3\rangle_1!}
\]
Here
\[
\langle n\rangle_1 = 1+q_1+q_1^2 + \ldots + q_1^{n-1}
\]
In this table $f_3$ and $f_0$ are polynomials in $\bar bb$
that may be computed by (11.1)-(11.3).

\vskip.5cm

\section{Lepton Neutrino Couplings$^{11}$}

Let us consider the process
\be
\bar\ell(j) + W^-\to \bar\nu(i) \quad \mbox{or} \quad
W^- \to \ell(j) + \bar\nu(i)
\ee
The standard matrix element for the absorption of a $W^-$
boson by a $\bar\ell(j)$ with the emission of a $\bar\nu(i)$
is by (10.6) to be multiplied by the following factor
\be
m(i,j) = \langle i|\stackrel{=}{D}^{3/2}_{-\frac{3}{2}\frac{3}{2}}
D^3_{30}\bar D^{3/2}_{\frac{3}{2}\frac{3}{2}}|j\rangle
\ee
where the double bar signifies an antiparticle in the final state.
By the Tables (11.1) and (11.2)
\be
m(i,j) = \Delta^3_{30}\left\langle\matrix{6\cr 3}\right\rangle_1
q_1^3\langle i|(-\bar b^3)(a^3b^3)(\bar a^3)|j\rangle
\ee
or
\be
m(i,j) = m(n)\delta(i,j)
\ee
where
\be
m(n) = -\left\langle\matrix{6\cr 3}\right\rangle_1^{1/2}
q^{6+6n}|\beta|^6 f(n)f(n+1)f(n+2)
\ee
and
\be
n=n_i = n_j
\ee
Here
\be
f(n) = 1-q^{2n}|\beta|^2
\ee
The form factor $m(i,j)$ vanishes if $i\not= j$ 
and depends on $n$ as shown.  Then define
\be
R_n \equiv \frac{m(n+1)}{m(n)} = q^6
\frac{1-q^{2n+6}|\beta|^2}{1-q^{2n}|\beta|^2}
\ee
including
\be
R_0 = q^6 \frac{1-q^6|\beta|^2}{1-|\beta|^2}
\ee
and
\be
R_1 = q^6 \frac{1-q^8|\beta|^2}{1-q^2|\beta|^2}
\ee
or by (12.9)
\be
|\beta|^2 = \frac{R_0-q^6}{R_0-q^{12}}
\ee
and by (12.10)
\be
|\beta|^2 = q^{-2} \frac{R_1-q^6}{R_1-q^{12}}
\ee
The universal Fermi interaction requires
\be
R_0 = R_1=1
\ee
and implies by (12.11) and (12.12)
\be
\begin{array}{rcl}
q &=& 1 \\
|\beta| &=& \frac{\sqrt{2}}{2} = .707
\end{array}
\ee
If the universal Fermi interaction is not exactly satisfied,
then the values of $(q,\beta)$ as determined by $R_0$ and $R_1$
will be shifted slightly away from (12.14).

\vskip.5cm

\section{Charge Changing Quark Couplings$^{11}$}

Let $Q\left(-\frac{1}{3},j\right)$ be any quark of charge
$-\frac{1}{3}e$ and let $Q\left(\frac{2}{3},j\right)$ be any quark of charge $+\frac{2}{3}e$.
Then we consider
\be
Q\left(-\frac{1}{3},j\right) + W^+ \to Q\left(\frac{2}{3},i
\right)
\ee
and denote its form factor by
\be
\langle i|\bar D^{3/2}_{-\frac{3}{2}-\frac{1}{2}}
D^3_{-30} D^{3/2}_{\frac{3}{2}-\frac{1}{2}}|j\rangle =
m(n)\delta(n,n_i)\delta(n,n_j)
\ee
where
\be
m(n) = -Cq^{6n+4}|\beta|^6 f(n)f(n+1)f(n-1)
\ee
Here the quark states are denoted by
\be
D^{3/2}_{\frac{3}{2}-\frac{1}{2}}|i\rangle \quad \mbox{and}\quad
D^{3/2}_{-\frac{3}{2}-\frac{1}{2}}|i\rangle
\ee
and the corresponding state of $W^+$ by
\[
D^3_{-30}
\]
In (13.3)
\be
C = \Delta^{3/2}_{\frac{3}{2}-\frac{1}{2}}
\Delta^{3/2}_{-\frac{3}{2}-\frac{1}{2}}
\left(\left\langle\matrix{3\cr 1}\right\rangle_1\right)^2\cdot
\Delta^3_{-30}\left\langle\matrix{6\cr 3\cr}\right\rangle_1
\ee
and $\Delta^i_{mm^\prime}$ is given by (11.2).
Again defining $R_n$ by
\be
R_n = \frac{m(n+1)}{m(n)}
\ee
we have
\begin{eqnarray}
R_0 &=& q^6 \frac{1-q^4|\beta|^2}{1-q^{-2}|\beta|^2} =
\frac{m(s+W^+\to c)}{m(d+W^+\to u)}  \\
R_1 &=& q^6\frac{1-q^6|\beta|^2}{1-|\beta|^2} =
\frac{m(b+W^+\to t)}{m(s+W^+\to c)}
\end{eqnarray}
Then
\be
|\beta|^2 = q^2\frac{R_0-q^6}{R_0-q^{12}} \quad \mbox{and}
\quad |\beta|^2 = \frac{R_1-q^6}{R_1-q^{12}}
\ee
We again have
\be
\begin{array}{rcl}
q &=& 1 \\
|\beta| &=& \frac{1}{2} \sqrt{2} = .707
\end{array}
\ee
if $R_0=R_1=1$ but according to the Kobayashi-Maskawa matrix we have
\begin{eqnarray}
R_0 &=& \left(\frac{.973}{.974}\right)^{1/2} = .999 \\
R_1 &=& \left(\frac{.999}{.973}\right) = 1.01
\end{eqnarray}

Since the diagonal elements of the Kobayashi-Maskawa matrix
are not quite equal, $q$ and $|\beta|$ differ slightly from
$q=1,|\beta|=.707$, if these Kobayashi-Maskawa ratios are
attributed entirely to the knot form factors.

\vskip.3cm

\begin{center}
\bf {Table 13.1: The Kobayashi-Maskawa (KM) Matrix}
\end{center}
\[
\begin{array}{c|ccc}
& d & s & b \\
\hline
u & 0.974 & 0.226 & 0.00359 \\
c & 0.226 & 0.973 & 0.0415 \\
t & 0.009 & 0.0407 & 0.999 \\
\end{array}
\]

The elements (13.2) are taken between states $|b^\prime\rangle$ that
are eigenstates of the Hamiltonian $H(\bar b,b)$ and therefore may
be regarded as states of definite mass.  One may similarly regard
the eigenstates of $d$ and $a$, that are raising and lowering, or
creation and annihilation operators, as flavor states, which are
superpositions of mass states.  We have seen that matrix elements 
like (13.2), describing transitions between quarks that are mediated
by weak bosons, are diagonal when taken between the mass eigenstates
$|b^\prime\rangle$.  They are not diagonal, however, in accord with
Table (13.1), when taken between flavor states as follows:
\[
\langle d^\prime|M|a^\prime\rangle =
\sum_{b^\prime b^{\prime\prime}} \langle d^\prime|b^\prime\rangle
\langle b^\prime|M|b^{\prime\prime}\rangle\langle b^{\prime\prime}|
a^\prime\rangle
\]
If one requires that $\langle d^\prime|M|a^\prime\rangle$, taken
between flavor states, be a $SU(3)$ matrix, then it is natural to
parametrize by $q,|\beta|$, and the three complex eigenvalues of 
$a$.$^{19}$

\vskip.5cm

\section{The Preon Representation$^{15,16}$}

The model so far described is based on representations
$D^{N/2}_{\frac{w}{2}\frac{r+1}{2}}$ of the knot algebra
$SL_q(2)$ or
$SU_q(2)$, where the integers $(N,w,r)$ label classical knots.
With these representations we define 
$D^{N/2}_{\frac{w}{2}\frac{r+1}{2}}|n\rangle$ to be the state
of a ``quantum knot" where $D^{N/2}_{\frac{w}{2}\frac{r+1}{2}}
(a,b,c,d)$ may be regarded as a kinematic factor. 
This state is additionally an eigenstate of
the Hamiltonian that defines the dynamics
of the knot.

Since the empirical basis of these considerations is the 
correspondence between the 4 quantum trefoils and the 4
families of fermions, we have focused on trefoils $(N=3)$ and
therefore on $j=\frac{N}{2}=\frac{3}{2}$.  As we have seen
the 4 families (neutrinos, leptons, up quarks, down quarks) may
all be represented by elements of the 3/2 representation
of $SL_q(2)$.  They
can also all be represented by $D^{3t}_{-3t_3-3t_0}$
corresponding to the fact that 
the isotopic spin $t=\frac{1}{2}$ for all the
elementary fermions.

It is then natural to examine the adjoint $(j=1)$ and 
fundamental $(j=1/2)$ representations.  To do this we extend
the results found for the 3/2 representation, i.e. we set
\begin{eqnarray}‫
j &=& \frac{N}{2} = 3t \\
m &=& \frac{w}{2} = -3t_3 \\
m^\prime &=& \frac{r+1}{2} = -3t_0 \\
Q &=& -\frac{e}{3} (m+m^\prime) = -\frac{e}{3} 
(w+r+1) = e(t_3+t_0)
\end{eqnarray}
According to Eqs. (11.1)-(11.3) and ignoring numerical
factors, the fundamental and adjoint representations are shown
in Tables (14.1) and (14.2).

\begin{center}
{\bf Table 14.1} \hspace{2.0in}   {\bf Table 14.2}
\end{center}
%\vskip.3cm
\[
D^{1/2}_{mm^\prime}: \quad  
\begin{array}{c|cc}
{}_m\backslash{}^{m'}
 & \frac{1}{2} & -\frac{1}{2} \\ \hline 
\frac{1}{2} & a & b \\
-\frac{1}{2} & c & d
\end{array} \hspace{1.5cm} 
D^1_{mm^\prime}: \quad  
\begin{array}{c|ccc}
{}_m\backslash{}^{m'}
 & 1 & 0 & -1 \\ \hline 
1 & a^2 & ab & b^2 \\
0 & ac & ad+bc & bd \\
-1 & c^2 & cd & d^2
\end{array} 
\]
According to (14.2)-(14.4) one finds the values of $(t_3,t_0,Q)$
shown in Tables (14.3) and (14.4)
\vskip.3cm

\begin{center}
{\bf Table 14.3}
\end{center}
\[
\begin{array}{c|ccc}
 & \underline{t_3} & \underline{t_0} & \underline{Q}  \\ 
a & -\frac{1}{6} & -\frac{1}{6} & -\frac{e}{3} \\
c & \frac{1}{6} & -\frac{1}{6} & 0 \\
d & \frac{1}{6} & \frac{1}{6} & \frac{e}{3} \\
b & -\frac{1}{6} & \frac{1}{6} & 0
\end{array}
\]
\vskip.5cm

\begin{center}
{\bf Table 14.4}
\end{center}

\vskip.3cm

\begin{table}[h]

\begin{center}
\begin{tabular}{l|cccc||l|cccc||l|cccc}
& \underline{$t_3$} & \underline{$t_0$} & \underline{$Q/e$} & 
\underline{$D^1_{mm^\prime}$} & &
\underline{$t_3$} & \underline{$t_0$} & \underline{$Q/e$} &
\underline{$D^1_{mm^\prime}$} & &
\underline{$t_3$} & \underline{$t_0$} & \underline{$Q/e$} &
\underline{$D^1_{mm^\prime}$} \\
$D^1_{11}$ & $-\frac{1}{3}$ & $-\frac{1}{3}$ &
$-\frac{2}{3}$ & $a^2$ & $D^1_{01}$ & 0 &
$-\frac{1}{3}$ & $-\frac{1}{3}$ & $ac$ & 
$D^1_{-11}$ & $\frac{1}{3}$ & $-\frac{1}{3}$ & 0 &
$c^2$ \\
$D^1_{10}$ & $-\frac{1}{3}$ & 0 & $-\frac{1}{3}$ &
$ab$ & $D^1_{00}$ & 0 & 0 & 0 & $ad+bc$ &
$D^1_{-10}$ & $\frac{1}{3}$ & 0 & $\frac{1}{3}$ &
$cd$ \\
$D^1_{1-1}$ & $-\frac{1}{3}$ & $\frac{1}{3}$ & 0 &
$b^2$ & $D^1_{0-1}$ & 0 & $\frac{1}{3}$ &
$\frac{1}{3}$ & $bd$ & $D^1_{-1-1}$ & $\frac{1}{3}$
& $\frac{1}{3}$ & $\frac{2}{3}$ & $d^2$
\end{tabular}
\end{center}
\end{table}

In these tables $a$ and $d$ have opposite values of the charge and hypercharge, while $c$ and $b$ are both neutral with opposite
values of $t_3$ and $t_0$.

We next relate the fundamental and adjoint representations to
the knot representations shown in Table 14.5.

\begin{center}
{\bf Table 14.5}
\end{center}
\[
\begin{array}{ccccc}
\underline{\mbox{Elementary Fermions}} &
\underline{D^{N/2}_{\frac{w}{2}\frac{r+1}{2}}} &
\underline{D^{3/2}_{\frac{w}{2}\frac{r+1}{2}}} &
\underline{Q_e} & \underline{et_0}  \\
(e^-,\mu^-,\tau^-) &
D^{3/2}_{\frac{3}{2}\frac{3}{2}} & a^3 & -e &
-\frac{e}{2} \\
(\nu_e,\nu_\mu,\nu_\tau) &
D^{3/2}_{-\frac{3}{2}\frac{3}{2}} & c^3 & 0 
& -\frac{e}{2} \\
(d,s,b) &
D^{3/2}_{\frac{3}{2}-\frac{1}{2}} & \sim ab^2 &
-\frac{1}{3} e & \frac{e}{6} \\
(u,c,t) &
D^{3/2}_{-\frac{3}{2}-\frac{1}{2}} & \sim cd^2 &
\frac{2}{3} e  & \frac{e}{6} \\
\end{array}
\]

In Table 14.5 the four monomials in the knot algebra represent
the four fermionic knots.  This table may also be interpreted by
regarding the element $a$ as a creation operator for a preon of
charge $-e/3$, and hypercharge $-e/6$ and by regarding $d$ as
a creation operator for a preon of charge $+e/3$ and hypercharge
$+e/6$ while $b$ and $c$ are regarded as creation operators for
neutral preons with hypercharge $e/6$ and $-e/6$ respectively.
This interpretation is consistent with the charge and 
hypercharge assignments in Tables 14.3 and 14.4 and also with
our conclusion from earlier work that adjoint operators
$(a,d)$ correspond to opposite charges and that the $(b,c)$
sector describes neutral states.  According to the same picture
the fermion knots, like the nucleons, are composed of three
fermions, which are now preons.

The results shown in these tables illustrate the following
general statements that can be proved for any knot described
by $D^{3t}_{-3t_3-3t_0}(a,b,c,d)$:
\begin{eqnarray}
t_3 &=& -\frac{1}{6} (n_a-n_d+n_b-n_c) \\
t_0 &=& -\frac{1}{6} (n_a-n_d-n_b+n_c) \\
Q &=& \frac{e}{3} (n_d-n_a)
\end{eqnarray}
where $(n_a,n_b,n_c,n_d)$ are the exponents of $(a,b,c,d)$
respectively in (5.19).  Since the $(a,b,c,d)$ are now
interpreted as creation operators for $(a,b,c,d)$ preons, the
$(n_a,n_b,n_c,n_d)$ are now the numbers of $(a,b,c,d)$ 
particles.  The $(n_a,n_b,n_c,n_d)$ vary among the terms
contributing to $D^{N/2}_{\frac{w}{2}\frac{r+1}{2}}$ but
$n_a-n_d$, $n_b-n_c$, and $n_a+n_b+n_c+n_d$ are the same for 
every term, and therefore characterize
$D^{N/2}_{\frac{w}{2}\frac{r+1}{2}} = D^{3t}_{-3t_3-3t_0}$.

In addition to (14.5)-(14.7) one has
\be
\begin{array}{rcl}
N^\prime &\equiv& n_a+n_b+n_c+n_d \\
&=& 2j = N
\end{array}
\ee
where $N^\prime$ is the total number of preons in the knot.
{\bf Therefore the total number of preons $(N^\prime)$ 
in the knot described by
$D^{N/2}_{\frac{w}{2}\frac{r+1}{2}}$ is equal to the number of
crossings $(N)$.}  Since the preons are to be regarded as
fermions, $D^{N/2}_{\frac{w}{2}\frac{r+1}{2}}$ represents a
boson or a fermion depending on whether the number of
crossings is even or odd, as we have previously assumed.

The relations (14.5)-(14.7) may be shown as follows.  By
(8.17)-(8.19) the operators $({\cal{N}},{\cal{W}},{\cal{R}})$
have eigenvalues $(2j,2m,2m^\prime)$ and if one extends
(9.19)-(9.21) to all representations one has
\be
(2j,2m,2m^\prime) = (6t,-6t_3,-6t_0) = (N,w,r+1)
\ee
On the other hand, the eigenvalues of $({\cal{N}},{\cal{W}},
{\cal{R}})$ when these operators are defined by (8.13)-(8.16),
may also be expressed as
\be
(n_a+n_b+n_c+n_d;n_a-n_d+n_b-n_c;n_a-n_d-n_b+n_c)
\ee
Then (14.9) and (14.10) imply (14.5)-(14.7) and (14.8).

If we maintain the relation $j=N/2$, then $j=1/2$ and $j=1$ imply
values of $N<3$.  Since the minimum value of $N$ is three for
a classical knot, $j=1/2$ and $j=1$ do not qualify as images of
classical knots.  They may be pictured as quantum 
images of twisted loops.  Viewed
as a particle a fermion becomes a boson by emitting or absorbing
a preon.  Viewed as a fermionic knot it becomes a bosonic knot by
adding or subtracting a curl.  A curl in turn is a twisted loop that
has been cut.
  
Although our definition of the quantum knot is
based on the knot algebra, it does not follow that the quantum
knot closely resembles the geometrical knot, just as the
quantum harmonic oscillator does not closely resemble the
classical harmonic oscillator.  In particular, although the
fundamental $D^{1/2}$ and adjoint $D^1$ representations
have $N<3$ and therefore do not qualify as images of knots in
the classical sense, they are not thereby disqualified as 
physical just as zero-point oscillations of the harmonic
oscillator are not physically disqualified.  Therefore we shall
take the view that the $D^{1/2}_{mm^\prime}$ fermions and the
$D^1_{mm^\prime}$ bosons qualify as particles of new fields and
we shall assume that they are subject to the Lagrangian of the
standard theory and may be discussed in the same way as the fermions and bosons of standard theory.

\vskip.5cm

\section{Preons as Physical Particles$^{15}$}

We shall now no longer regard the preons as merely 
a simple way to describe
the algebraic structure of the knot polynomials.  If these preons
are in fact physical particles, the following decay modes of the
quarks are possible
\[
\begin{array}{llll}
{\rm down~quarks:} & D^{3/2}_{\frac{3}{2}-\frac{1}{2}}
\longrightarrow D^{1/2}_{\frac{1}{2}\frac{1}{2}} +
D^1_{1-1} & & ab^2 \longrightarrow a+b^2 \\
& & {\rm or} & \\
{\rm up~quarks:} & D^{3/2}_{-\frac{3}{2}-\frac{1}{2}}
\longrightarrow D^{1/2}_{-\frac{1}{2}\frac{1}{2}} +
D^1_{-1-1} & & cd^2 \longrightarrow c+d^2
\end{array}
\]
and the preons could play an intermediary role as virtual
particles in quark processes.  

The justification for considering
the preons seriously as physical particles would then no longer 
depend on the knot
conjecture but rather on a more general role of $SL_q(2)$ gauge
invariance.  Then the preons would appear as matrix elements of
the fundamental and adjoint representations of $SL_q(2)$ just as
the fermionic and bosonic 
quantum knots appear in the $j=3/2$ and $j=3$
representations of $SL_q(2)$.$^3$  In this scenario 
quantum knots would be
just one of the manifestations of a $SL_q(2)$ 
related symmetry.
There would also be no need to introduce a new Lagrangian for the
preons since all particles described by
representations of $SL_q(2)$ would be subject to
the same modified standard action.

\vskip.5cm

The simple knot model predicts an unlimited number of excited
states$^{2,3}$ but it appears that there are only three 
generations, e.g.
$(d,s,b)$.  According to the preon scenario, however, it may be
possible to avoid this problem by showing
that the quarks will dissociate into preons if
given a critical ``dissociation energy" less than that needed to
reach the level of the fourth predicted flavor.  In that case one
would also
expect the formation of a preon-quark plasma at sufficiently
high temperatures.  It may be possible to study the thermodynamics of the  plasma composed of quarks and these hypothetical particles.

Since the $a$ and $\bar a$ particles are charged $(\pm e/3)$
one should also expect their electroproduction according to
\[
e^++e^- \rightarrow a+\bar a + \ldots
\]
at sufficiently high energies of a colliding $(e^+,e^-)$ pair.
More generally one should expect
\[
? \to \gamma \to a + \bar a + \ldots
\]
if $\gamma$ is sufficiently energetic independent of how it is
produced.

\vskip.5cm

\section{Field Theory of Quantum Knots$^{16}$}

Let us introduce the field $\Psi^j_{mm^\prime}(x;abcd)$, the
product of 
the standard point particle field, $\psi^j_{mm^\prime}(x)$,
and an internal factor $D^j_{mm^\prime}$, as follows:
\be
\Psi^j_{mm^\prime} = \psi^j_{mm^\prime}(x)
D^j_{mm^\prime}(a,b,c,d)
\ee
These fields undergo transformations of the Poincar\'e
algebra when the spacetime points $(x)$ are relabelled 
by transformations that
preserve the structure of spacetime, and they also undergo
gauge transformations when the discrete elements $(a,b,c,d)$ are
transformed so as to preserve the structure of the knot algebra.
By the usual argument the Lagrangian must be constructed to be
invariant under 
all of these transformations since the 
relabelling of the continuum and the algebra must not influence
the physics.  There is then, by Noether's theorem, a conserved
quantity for each independent gauge transformation.  In the
familiar way the eigenvalues of the corresponding conserved
and commuting operators are used to label the particles and some
of these eigenvalues are functions of the 
$(j,m,m^\prime)$ that label the knot particles.

In much of the following discussion we do not distinguish between
$SL_q(2)$ 
and $SU_q(2)$ but in part of the work we may explicitly refer to
$SU_q(2)$ by setting $d=\bar a$ and $c=-q_1\bar b$.

In the knot electroweak theory, just as in the standard model,
the fields representing the fermion families may be arranged
by (16.1) in two isotopic doublets as follows:
\be
\begin{array}{rl}
{\rm leptons:} & \Psi_\ell = \psi_\ell(x) D^{3/2}_{\frac{3}{2}
\frac{3}{2}}(abcd) \\
{\rm neutrinos:} & \Psi_\nu = \psi_\nu(x) D^{3/2}_{-\frac{3}{2}
\frac{3}{2}}(abcd)
\end{array}
\ee
and
\be
\begin{array}{rl}
{\rm down~ quarks:} & \Psi_d = \psi_d(x)D^{3/2}_{\frac{3}{2}
\frac{1}{2}}(abcd) \\
{\rm up~quarks:} & \Psi_u = \psi_u(x) D^{3/2}_{-\frac{3}{2}
\frac{1}{2}}(abcd)
\end{array}
\ee
Since the gauge transformations 
on the knot algebra induce corresponding gauge
transformations on the $D^j_{mm^\prime}$ according to (6.6b), we
then have in the lepton family
\be
\Psi^\prime_\ell = \psi_\ell(x)\left[e^{i\frac{3}{2}(\varphi_a+
\varphi_b)}e^{i\frac{3}{2}(\varphi_a-\varphi_b)}
D^{3/2}_{\frac{3}{2}\frac{3}{2}}(abcd)\right]
\ee
or
\be
\Psi^\prime_\ell = e^{i\frac{3}{2}(\varphi_a+\varphi_b)}
e^{i\frac{3}{2}(\varphi_a-\varphi_b)}\Psi_\ell
\ee
Likewise for the neutrino family we have
\be
\Psi^\prime_\nu = e^{-i\frac{3}{2}(\varphi_a+\varphi_b)}
e^{i\frac{3}{2}(\varphi_a-\varphi_b)}\Psi_\nu
\ee
For the quarks we have
\begin{eqnarray}
\Psi^\prime_d &=& e^{i\frac{3}{2}(\varphi_a+\varphi_b)}
e^{i\frac{1}{2}(\varphi_a-\varphi_b)}\Psi_d \\
\Psi^\prime_u &=& e^{-i\frac{3}{2}(\varphi_a+\varphi_b)}
e^{i\frac{1}{2}(\varphi_a-\varphi_b)}\Psi_u
\end{eqnarray}
The gauge transformations on the knot algebra therefore
induce the following diagonal $SU(2)$ transformations on the
$(\ell,\nu)$ and $(u,d)$ doublets
\be
\begin{array}{l}
\Psi_\ell^\prime = e^{i\frac{1}{2}\varphi_+}\Psi_\ell \\
\Psi^\prime_\nu = e^{-i\frac{1}{2}\varphi_+}\Psi_\nu
\end{array} \qquad
\begin{array}{l}
\Psi_d^\prime = e^{i\frac{1}{2}\varphi_+}\Psi_d \\
\Psi_u^\prime = e^{-i\frac{1}{2}\varphi_+}\Psi_u
\end{array}
\ee
as well as the following $U(1)$ transformations
\be
\begin{array}{l}
\Psi_\ell^\prime = e^{i\frac{1}{2}\varphi_-}\Psi_\ell \\
\Psi_\nu^\prime = e^{i\frac{1}{2}\varphi_-}\Psi_\nu
\end{array} \qquad
\begin{array}{l}
\Psi_d^\prime = e^{i\frac{1}{6}\varphi_-}\Psi_d \\
\Psi_u^\prime = e^{i\frac{1}{6}\varphi_-}\Psi_u
\end{array}
\ee
Here
\be
\begin{array}{rcl}
\varphi_+ &=& 3(\varphi_a+\varphi_b) \\
\varphi_- &=& 3(\varphi_a-\varphi_b)
\end{array}
\ee

In summary, the gauge transformations (G) on the algebra induce
diagonal $SU(2)\times U(1)$ transformations on the fermion
doublets as follows:
\be
\Psi^{3t~~~^\prime}_{-3t_3-3t_0} =
e^{-it_3\varphi_+}e^{-it_0\varphi_-}
\Psi^{3t}_{-3t_3-3t_0}
\ee
where
\be
t = \frac{1}{2},~t_3 = \pm\frac{1}{2},~t_0 = \frac{1}{2},
\frac{1}{6}
\ee

The same Eq. (16.12) holds for the vector boson triplet with
$t=1$ and the pair $(t_3,t_0)$ as recorded in Table 10.1.

The gauge transformations (16.12) referring to the $SL_q(2)$ or
$SU_q(2)$ are of course additional to the standard gauge transformations referring to isotopic $SU(2)$.

The statement (16.12) remains true if the transformations are local,
i.e., if
\be
\Psi^{3t~~~^\prime}_{-3t_3-3t_0} =
e^{-it_3\varphi_+(x)} e^{-it_0\varphi_-(x)}
\Psi^{3t}_{-3t_3-3t_0}
\ee
The preceding equation (16.14) permits the construction of an
Abelian field theory based solely on the knot algebra.

The gauge transformations $U_a\times U_b$ on the knot algebra
may also be written in doublet form as follows:
\begin{subequations}
\begin{eqnarray}
\left(
\begin{array}{c}
a \\ c
\end{array} \right)^\prime &=& \left(
\begin{array}{cc}
e^{\frac{1}{6}\varphi_+} & 0 \\
0 & e^{-\frac{1}{6}\varphi_+}
\end{array} \right) \left(
\begin{array}{cc}
e^{\frac{1}{6}\varphi_-} & 0 \\
0 & e^{\frac{1}{6}\varphi_-}
\end{array} \right) \left(
\begin{array}{c} 
a \\ c
\end{array} \right) \\
\left(
\begin{array}{c}
b \\ d
\end{array} \right)^\prime &=& \left(
\begin{array}{cc}
e^{\frac{i}{6}\varphi_+} & 0 \\
0 & e^{-\frac{i}{6}\varphi_+}
\end{array} \right) \left(
\begin{array}{cc}
e^{-\frac{i}{6}\varphi_-} & 0 \\
0 & e^{-\frac{i}{6}\varphi_-}
\end{array} \right) \left(
\begin{array}{c}
b \\ d
\end{array} \right)
\end{eqnarray}
\end{subequations}
where the components of the two doublets are preons.

Here
\begin{subequations}
\begin{eqnarray}
\varphi_a &=& \frac{1}{6} (\varphi_++\varphi_-) \\
\varphi_b &=& \frac{1}{6}(\varphi_+-\varphi_-)
\end{eqnarray}
\end{subequations}
Transformations that mix $a$ and $c$ or $b$ and $d$, however,
do not leave the knot algebra (A) invariant.  Hence (16.14)
cannot be extended to off-diagonal $SU(2)$ transformations,
i.e. to
\be
\Psi^{3t~~~^\prime}_{-3t_3-3t_0} = e^{-i\vec t\vec\varphi_+(x)}
e^{-it_0\varphi_-(x)} \Psi^{3t}_{-3t_3-3t_0}
\ee
and therefore a non-Abelian field theory cannot be
supported solely by the gauge group of the knot algebra.  On
the other hand, the isotopic spin $\times$ hypercharge group,
$SU(2)\times U(1)$, is empirically required and the standard electroweak theory postulates that this group is local.

If the vector field is 
introduced as the connection of this local
group in the standard way, then in the $q$-knot modification 
of the standard theory, one may represent the
vector fields by (16.1) where $\psi^j_{mm^\prime}$ transforms 
under local 
$SU(2)\times U(1)$ and where the second factor in (6.1), $D^j_{mm^\prime}$,
transforms according to the global gauge symmetry of the knot
algebra. The two symmetries, local $SU(2)\times U(1)$ and global
$U(a)\times U(b)$, are matched by requiring $(j,m,m^\prime) =
3(t_1-t_3,-t_0)$. 

\vskip.5cm

\section{Vector Fields and their Field Strengths}

In the standard theory the vector bosons are quanta of the vector
fields and the vector fields are connections of an underlying
gauge group.  In the standard electroweak model there are two
gauge groups, namely $SU(2)$, the isotopic spin group, and
$U(1)$, the hypercharge group, and the corresponding vector
connection of $SU(2)\times U(1)$ is
\be
W_+t_+ + W_-t_- + W_3t_3 + W_0t_0
\ee
where the $t_k$ are the generators of the Lie algebras of
$SU(2)$ and $U(1)$:
\be
t_+ = \left(
\begin{array}{cc}
0 & 1 \\ 0 & 0
\end{array} \right) \quad t_- = \left(
\begin{array}{cc}
0 & 0 \\ 1 & 0
\end{array} \right) \quad t_3 = \left(
\begin{array}{cc}
1 & 0 \\ 0 & -1
\end{array} \right) \quad t_0 = \left(
\begin{array}{cc}
1 & 0 \\ 0 & 1
\end{array} \right)
\ee
By Eqs. (16.1) and (17.1) the 
corresponding connection in the knot model is
\begin{eqnarray}
& & (W_+t_+) {\cal{D}}_++(W_-t_-)
{\cal{D}}_- + (W_3t_3){\cal{D}}_3+(W_0t_0){\cal{D}}_0 \\
&=& W_+(t_+{\cal{D}}_+) + 
W_-(t_-{\cal{D}}_-) + W_3(t_3{\cal{D}}_3) + W_0(t_0{\cal{D}}_0)
\end{eqnarray}
where the knot factors are
\[
({\cal{D}}_+,{\cal{D}}_-,{\cal{D}}_3,{\cal{D}}_0) \sim 
(D^3_{-30},D^3_{30},D^3_{00},D^0_{00})
\]
and the $D^j_{mm^\prime}$ are given in Table 10.1.  We are now 
referring explicitly to $SU_q(2)$.

Let us therefore define
\be
{\cal{W}}_\mu = ig\vec W_\mu\vec\tau + ig_0W_\mu^0\tau_0
\ee
where
\begin{eqnarray}
\tau_\pm &\equiv& c_\pm t_\pm {\cal{D}}_\pm \\
\tau_3 &\equiv& c_3 t_3 {\cal{D}}_3 \\
\tau_0 &\equiv& c_0 t_0 {\cal{D}}_0
\end{eqnarray}
and
\begin{eqnarray}
{\cal{D}}_+ &=& \bar b^3\bar a^3 \\
{\cal{D}}_- &=& a^3b^3 \\
{\cal{D}}_3 &=& f(b\bar b) \\
{\cal{D}}_0 &=& 1
\end{eqnarray}
Here the ${\cal{D}}_k~(k=+,-,3)$ differ from the 
$D^{3t}_{-3t_3-3t_0}$ only by factors
absorbed in the $c_k$.

The $c_k$ will be determined in Section 19.
 
Let us define by (17.5) a covariant derivative
\be
\nabla_\mu \equiv \partial_\mu + {\cal{W}}_\mu
\ee
that satisfies
\be
\nabla_\mu^\prime = S\nabla_\mu S^{-1}
\ee
where $S\epsilon SU(2)\times U(1)\times U_a(1)\times
U_b(1)$.  Then
\be
{\cal{W}}^\prime_\mu = S{\cal{W}}_\mu S^{-1} + S\partial_\mu
S^{-1}
\ee
The corresponding field strengths are
\be
{\cal{W}}_{\mu\lambda} = (\nabla_\mu,\nabla_\lambda)
\ee
that transform as
\be
{\cal{W}}_{\mu\lambda}^\prime = S{\cal{W}}_{\mu\lambda}
S^{-1}
\ee
We shall next ignore the $W_\mu^0$ vector field.
By (17.5), (17.13), and (17.16) the non-Abelian field strengths
are
\be
{\cal{W}}_{\mu\lambda} = ig(\partial_\mu W^m_\lambda-
\partial_\lambda W^m_\mu) \tau_m-g^2 W^m_\mu W^\ell_\lambda
[\tau_m,\tau_\ell]
\ee
Here
\be
\left[\tau_k,\tau_\ell\right] = c_kc_\ell\left[[t_k,t_\ell]
{\cal{D}}_k
{\cal{D}}_\ell + t_\ell t_k[{\cal{D}}_k,{\cal{D}}_\ell]\right]
\ee
where
\begin{eqnarray}
\left[t_k,t_\ell\right] &=& c^s_{k\ell}t_s  \qquad
(k,\ell) = (+,-,3) \\
\left[{\cal{D}}_k,{\cal{D}}_\ell\right] &=& 
\hat c^s_{k\ell}{\cal{D}}_s \\
t_kt_\ell &=& \gamma^s_{k\ell}t_s + \frac{1}{2}
\delta(k,\pm)\delta(\ell,\mp) \\
{\cal{D}}_k{\cal{D}}_\ell &=& \hat\gamma^s_{k\ell}
{\cal{D}}_s
\end{eqnarray}
Then
\be
\left[\tau_k,\tau_\ell\right] = \frac{c_kc_\ell}{c_s}
C^s_{k\ell}\tau_s + \frac{1}{2} c_kc_\ell \delta(k,\pm)
\delta(\ell,\mp) \hat c^s_{k\ell}{\cal{D}}_s
\ee
where
\be
C^s_{k\ell} = c^s_{k\ell}\hat\gamma^s_{k\ell} +
\gamma^s_{\ell k}\hat c^s_{k\ell}
\ee
The structure coefficients of these algebras, including
$c^s_{k\ell}$ and $\gamma^s_{\ell k}$ as
well as the $\hat c^s_{k\ell}$ and $\hat\gamma^s_{k\ell}$,
commute since they are either numerically valued or 
are functions of $\bar bb$. They are all
numerically valued when allowed to operate on states $|n\rangle$ of the $q$-oscillator.

It follows from (17.18) 
and (17.24) that the field strengths are given by
\be
{\cal{W}}_{\mu\lambda} = W^s_{\mu\lambda}\tau_s + 
\hat W^s_{\mu\lambda}{\cal{D}}_s
\ee
where
\be
W^s_{\mu\lambda} = ig(\partial_\mu W^s_\lambda - \partial_\lambda
W^s_\mu) - g^2c_mc_\ell c_s^{-1} C^s_{m\ell}W_\mu^m W^\ell_\lambda
\ee
and
\be
\hat W^s_{\mu\lambda} = -\frac{1}{2} g^2 c_mc_\ell
\delta(\ell,\pm)\delta(m,\mp) \hat c^s_{m\ell} W_\mu^m
W^\ell_\lambda
\ee

\vskip.5cm

\section{Interactions of the Vector Fields}

\no (a) Self-Interactions.

We choose as the vector field invariant the following expectation
value:
\be
I = \langle 0|{\rm Tr}~{\cal{W}}_{\mu\lambda}{\cal{W}}^{\mu
\lambda}|0\rangle
\ee
where $|0\rangle$ is the ground state of the $q$-oscillator
defined in Section 7.  Here the trace is taken over the part
dependent on the $t_k$.

To reduce I consider
\be
\begin{array}{rcl}
\langle 0|{\cal{W}}_{\mu\lambda}{\cal{W}}^{\mu\lambda}|0\rangle
&=& {\displaystyle\sum_n} \{\langle 0|W^s_{\mu\lambda}W^{r\mu\lambda}|n\rangle
\langle n|\tau_s\tau_r|0\rangle \\
& &{} +\langle 0|\hat W^s_{\mu\lambda}\hat W^{r\mu\lambda}
|n\rangle\langle n|{\cal{D}}_s{\cal{D}}_r|0\rangle \\
& &{} + \langle 0|W^s_{\mu\lambda}\hat W^{r\mu\lambda}|n\rangle
\langle n|\tau_s{\cal{D}}_r|0\rangle \\
& &{} + \langle 0|\hat W^r_{\mu\lambda}W^{s\mu\lambda}
|n\rangle\langle n|{\cal{D}}_r\tau_s|0\rangle\}
\end{array}
\ee
where the sum is over all states of the $q$-oscillator.

Since $W^s_{\mu\lambda}$ and $\hat W^s_{\mu\lambda}$ depend on
the algebra 
of $SU_q(2)$ only through $\bar bb$, they have no off-diagonal
elements in $n$.  Then
\be
\begin{array}{rcl}
{\rm Tr}\langle 0|{\cal{W}}_{\mu\lambda}{\cal{W}}^{\mu\lambda}|0\rangle
&=&{\rm Tr}\{\langle 0|W^s_{\mu\lambda}W^{r\mu\lambda}|0\rangle
\langle 0|\tau_s\tau_r|0\rangle \\
& & +\langle 0|\hat W^s_{\mu\lambda}\hat W^{r\mu\lambda}
|0\rangle\langle 0|{\cal{D}}_s{\cal{D}}_r|0\rangle
+ \langle 0|W^s_{\mu\lambda}\hat W^{r\mu\lambda}|0\rangle \\
& & \times \langle 0|\tau_s{\cal{D}}_r|0\rangle +
\langle 0|\hat W^r_{\mu\lambda}W^{s\mu\lambda}|0\rangle
\langle 0|{\cal{D}}_r\tau_s|0\rangle\}
\end{array}
\ee 
To continue the reduction of (18.1), we next compute the
following factors in (18.3)
\be
\langle 0|{\rm Tr}~\tau_s\tau_r|0\rangle = c_sc_r({\rm Tr}~
t_st_r)\langle 0|{\cal{D}}_s{\cal{D}}_r|0\rangle
\ee
where
\begin{eqnarray}
& &\langle 0|{\cal{D}}_s{\cal{D}}_r|0\rangle =
[\delta(s,\pm)\delta(r,\mp) + \delta(s,3)\delta(r,3)]
\langle 0|\bar{\cal{D}}_r{\cal{D}}_r|0\rangle \\
& &\langle 0|{\rm Tr}~\tau_s{\cal{D}}_r|0\rangle =
\langle 0|{\rm Tr}~{\cal{D}}_r\tau_s|0\rangle = 0
\end{eqnarray}
Then the field invariant reduces to
\be
I = \sum_{s,r=(+,-)}\langle 0|A_{sr}W^s_{\mu\lambda}
W^{r\mu\lambda} + 2\hat W^s_{\mu\lambda}\hat W^{r\mu\lambda}
|0\rangle\langle 0|{\cal{D}}_s{\cal{D}}_r|0\rangle
\ee
where
\be
A_{sr} = c_sc_r~{\rm Tr}~t_st_r
\ee
Here $W^s_{\mu\lambda}$ and $\hat W^s_{\mu\lambda}$ are given
by (17.27) and (17.28) and the matrix elements
$\langle 0|{\cal{D}}_s{\cal{D}}_r|0\rangle = 0$ unless
${\cal{D}}_s = \bar{\cal{D}}_r$.  Then
\begin{eqnarray}
& &\langle 0|\bar{\cal{D}}_+{\cal{D}}_+|0\rangle =
\langle 0|a^3b^3\bar b^3a^3|0\rangle = q^{18}
\langle 0|(\bar bb)^3a^3\bar a^3|0\rangle \\
& &\langle 0|\bar{\cal{D}}_-{\cal{D}}_-|0\rangle = 
\langle 0|\bar b^3\bar a^3a^3b^3|0\rangle =
\langle 0|\bar bb)^3\bar a^3a^3|0\rangle \\
& &\langle 0|\bar{\cal{D}}_3{\cal{D}}_3|0\rangle =
|f(\bar bb)|^2
\end{eqnarray}
where $f(\bar bb)$ is given by (17.11) and abbreviates $D^3_{00}$.
These matrix elements are all functions of $\bar bb$ since
\begin{eqnarray}
\bar a^na^n &=& \prod^n_{s=1}(1-q_1^{2s}\bar bb) \\
a^n\bar a^n &=& \prod^{n-1}_{s=0}(1-q^{2s}\bar bb)
\end{eqnarray}

The expression $W^s_{\mu\lambda}$ is of the same form as in the
standard theory but the structure coefficients differ from those
of the $SU(2)$ algebra because they depend on $\bar bb$.  Since
(18.7) is evaluated on the state $|0\rangle$ all expressions of
the form $F(\bar bb)$ become $F(|\beta|^2)$.  Therefore the
structure constants $C^s_{m\ell}(\bar bb)$ buried in
$W^s_{\mu\lambda}$ and in turn appearing in (18.7) become
$C^s_{m\ell}(|\beta|^2)$.  Then the final reduced form of
$\langle 0|{\rm Tr}~{\cal{W}}_{\mu\lambda}{\cal{W}}^{\mu\lambda}
|0\rangle$ in (18.7) will have one part $W^s_{\mu\lambda} W_s^{\mu\lambda}$
essentially the same as the standard theory,
but with structure constants $C^s_{m\ell}$ depending on
$|\beta|^2$.  There is also a second part $\hat W^s_{\mu\lambda}
\hat W_s^{\mu\lambda}$ depending on $\hat c^s_{m\ell}$ which is
dependent on $q$ and $\beta$.  The sum of these two parts is
multiplied by $\langle 0|\bar{\cal{D}}_s{\cal{D}}_s|0\rangle$,
also a function of $q$ and $\beta$.  
The functions
$c_s(q,\beta)$ will be given in Section 19.

\vskip.3cm

\no (b) Interactions of Vector Bosons and Fermions.

To describe the boson-fermion interaction we introduce
$\Psi_{Ari}$ where
\begin{eqnarray*}
\Psi_{1ri} &=& \psi_{1r} D_r(a,\bar a,b,\bar b)|i\rangle \qquad ~~~~
A = 1 \quad r = \nu,\ell~(i = 0,1,2) \\
\Psi_{2ri} &=& \psi_{2r} D_r(a,\bar a,b,\bar b)|i\rangle \qquad ~~~~
A = 2 \quad r = u,d~(i = 0,1,2)
\end{eqnarray*}
Then the boson-fermion interaction is contained in
\be
(\bar\Psi_A)_{ri}\nabla\!\!\!\!/_{rs}(\Psi_A)_{si^\prime}
\ee
where $A=1$ labels the $(\ell,\nu)$ doublet and $A=2$ labels the
quark doublet $(u,d)$.  To obtain improved agreement with 
experiment, with the Kobayashi-Maskawa matrix, and with neutrino
oscillations it is necessary to introduce flavor states as
described in Ref. 19.

In Ref. 16 it is shown that the entire action is invariant under the
$SU(2)\times U(1)\times U_a(1)\times U_b(1)$ symmetries.

\vskip.5cm

\section{The Higgs Sector in the Knot Model$^{16}$}

We follow the standard theory in discussing the vector masses and in
the process, we shall 
determine the constants $(c_\pm,c_3,c_0)$ introduced in (17.6)-(17.8).

The neutral couplings are by (17.5)
\be
i(gW_3\tau_3 + g_0W_0\tau_0)
\ee
Introducing the physical fields ($A$ and $Z$) in the standard way, we
have
\begin{eqnarray}
W_0 &=& A\cos\theta-Z\sin\theta \\
W_3 &=& A\sin\theta+Z\cos\theta
\end{eqnarray}
Then (19.1) expressing the neutral couplings becomes
\be
i({\cal{A}}A + {\cal{Z}}Z)
\ee
where
\begin{eqnarray}
{\cal{A}} &=& g\tau_3\sin\theta + g_0\tau_0\cos\theta \\
{\cal{Z}} &=& g\tau_3\cos\theta-g_0\tau_0\sin\theta
\end{eqnarray}
Now take $\theta$ to be the Weinberg angle.  Then
\be
\tan\theta = \frac{g_0}{g}
\ee
and by (19.5) and (19.6)
\begin{eqnarray}
{\cal{A}} &=& g_0(\tau_3+\tau_0)\cos\theta \\
{\cal{Z}} &=& g(\tau_3-\tau_0\tan^2\theta)\cos\theta
\end{eqnarray}
 Let $|\nu\rangle$ be any neutral state.  We shall require
\be
(\tau_3+\tau_0)|\nu\rangle = 0
\ee
where $|\nu\rangle$ is a numerically valued two component state.
Then by (19.8) and (19.9)
\begin{eqnarray}
{\cal{A}}|\nu\rangle &=& 0 \\
{\cal{Z}}|\nu\rangle &=& \frac{g}{\cos\theta} \tau_3|\nu\rangle
\end{eqnarray}
Hence the covariant derivative of a neutral state is
\be
\nabla = \partial + ig\left[W_+\tau_+ + W_-\tau_- +
\frac{Z\tau_3}{\cos\theta}\right]
\ee
Denote the neutral Higgs scalar (unitary gauge) by
\be
\phi = \rho(x) D_n|0\rangle
\ee
where $D_n$ is the internal state of the neutral Higgs.

Then the kinetic energy terms of the neutral Higgs in the standard
model is by (19.13) and (19.14)
\begin{eqnarray}
&&\frac{1}{2} {\rm Tr}{(\overline{\nabla_\mu\varphi}}\nabla^\mu\varphi)
\nonumber \\
&&= \frac{1}{2}~ {\rm Tr}\langle 0|\bar D_n  \nonumber \\
&& {}\times \left[\partial_\mu\rho\partial^\mu\rho + g^2\rho^2
[W_+^\mu W_{+\mu}\bar\tau_+\tau_+ + W^\mu_-W_{-\mu}\bar\tau_-\tau_-
+\frac{Z^\mu Z_\mu}{\cos^2\theta} \bar\tau_3\tau_3]\right] D_n|0\rangle \\
&&=I~\partial_\mu\rho\partial^\mu\rho+g^2\rho^2
\left[I_{++}W_+^\mu W_{+\mu} + I_{--}W_-^\mu W_{-\mu} +
\frac{I_{33}}{\cos^2\theta}~Z^\mu Z_\mu\right]
\end{eqnarray}
where
\be
\begin{array}{rcl}
I &=& \frac{1}{2}~{\rm Tr}\langle 0|\bar D_nD_n|0\rangle \\
I_{++} &=& \frac{1}{2}~{\rm Tr}\langle 0|\bar D_n\bar\tau_+\tau_+D_n
|0\rangle \\
I_{--} &=& \frac{1}{2}~{\rm Tr}\langle 0|\bar D_n\bar\tau_-\tau_-D_n
|0\rangle \\
I_{33} &=& \frac{1}{2}~{\rm Tr}\langle 0|\bar D_n\bar\tau_3\tau_3D_n
|0\rangle
\end{array}
\ee
To agree with the masses predicted by the standard model (19.16) must
be reduced to the following
\be
\partial_\mu\bar\rho\partial^\mu\bar\rho + g^2\bar\rho^2
\left[W_+^\mu W_{+\mu} + W^\mu_-W_{-\mu} + \frac{1}{\cos^2\theta}
Z^\mu Z_\mu\right]
\ee
where
\be
\bar\rho = I^{1/2}\rho
\ee
To achieve this reduction we impose the following relations
\be
\frac{I_{kk}}{I} = 1 \qquad k = (+,-,3)
\ee
or
\be
\frac{{\rm Tr}\langle 0|\bar D_n(\bar\tau_k\tau_k)D_n|0\rangle}
{{\rm Tr}\langle 0|\bar D_nD_n|0\rangle} = 1 \qquad k = (+,-,3)
\ee
By (17.6), (17.7) and (19.21) we have
\be
|c_k|^{-2} = \frac{\langle 0|\bar D_n(\bar {\cal{D}}_k
{\cal{D}}_k)D_n|0\rangle}
{\langle 0|\bar D_nD_n|0\rangle} \qquad k = (+,-,3)
\ee
In (17.6) and (17.7) the coefficients $(c_\pm,c_3)$ were introduced as
unknown factors.  They are now fixed by (19.22) as definite 
functions of $q$ and $\beta$.  Here the 
${\cal{D}}_k$ are given by (17.9)-(17.11).

The simplest assumption for the neutral Higgs is
\be
D_n = D^0_{00} = 1
\ee
Then
\be
|c_k|^{-2} = \langle 0| 
\bar{\cal{D}}_k{\cal{D}}_k|0\rangle \qquad k = (-,+,3) 
\ee
By (17.9)-(17.11)
\begin{eqnarray}
|c_-|^{-2} &=& \langle 0|\bar b^3\bar a^3 a^3b^3|0\rangle \\
&=& |\beta|^6 \prod^3_1 (1-q_1^{2t}|\beta|^2) \\
|c_+|^{-2} &=& \langle 0|a^3b^3\bar b^3\bar a^3|0\rangle \\
&=& q^{18}|\beta|^6 \prod^2_0 (1-q^{2t}|\beta|^2) \\
|c_3|^{-2} &=& \langle 0|\bar {\cal{D}}_3{\cal{D}}_3|0\rangle \\
&=& [f(|\beta|^2)]^2
\end{eqnarray}
If the $c_k$ satisfy the 
above relations, then the vector boson masses
satisfy the ratios of the standard theory according to (19.18).  With the
same assumption for the Higgs, we shall next compute the mass ratios
in each fermion family.

\vskip.5cm

\section{The Fermion Mass Term of the Standard Model$^{10}$}
  
In the knot model
there is a spectrum of masses that depends on the particular
Hamiltonian that is assumed for the knot.  We shall restrict this 
Hamiltonian by the requirements that it lies in the knot algebra and
that its eigenstates are $D^{N/2}_{\frac{w}{2}\frac{r+1}{2}}|
n\rangle$ as we have previously assumed, where 
$D^{N/2}_{\frac{w}{2}\frac{r+1}{2}}$ is the kinematic part
and the $|n\rangle$ are the eigenstates of the commuting $b$ and
$c$.  Finally, in order that the $H$ 
introduced in (8.6) qualify as the Hamiltonian of an
elementary fermionic knot we shall require that it be compatible
with the fermion mass term in the standard theory, namely
\be
{\cal{M}} = \bar L\varphi R + \bar R\varphi L
\ee 
where $L$ and $R$ are left- and
right-chiral Lorentz spinors and $\varphi$ is the Higgs field, a 
Lorentz scalar, so that product $\bar L\varphi R$ is Lorentz invariant.  
In the standard Lagrangian $L$ and $\varphi$ 
are isotopic doublets.
$(\bar L\varphi)$ and $R$ are separately 
isotopic singlets and ${\cal{M}}$
is invariant under the gauged $SU(2)\times U(1)$ group.

In the knot model $L$ is additionally a fermionic knot with the charge
structure $D^{3/2}_{-3t_3-3t_0}$. 
If a knot singlet is assigned to
$\varphi$, then $\varphi$ is neutral (unitary gauge) while the right
chiral spinor must have the same knot state as the left chiral
spinor, namely, $D^{N/2}_{\frac{w}{2}\frac{r+1}{2}}$, in order to
preserve the $U_a(1)\times U_b(1)$ invariance.  Then the standard
Higgs mechanism is still possible with $\varphi\sim D^0_{00}$.

One sees that if the knot state is 
$D^{3/2}_{\frac{w}{2}\frac{r+1}{2}}$ for both $L$ and $R$, the relation between $(t_3,t_0)$ and
$(w,r)$ is different for $L$ and $R$, but the expression for
charge, namely $-\frac{e}{6}(w+r+1)$ is the same for both.  In the
standard model $L$ and $R$ have different relations to the 
isotopic spin
group; here also they have different relations to 
the isotopic spin, but the 
same description in the knot algebra.

If $L$ and $R$ are now assigned the same internal state, and we treat
the mass term in the same way as the other terms of the Lagrangian,
then we have
\begin{eqnarray}
L &\to& \chi_L(w,r,n) D^{3/2}_{\frac{w}{2}\frac{r+1}{2}}|n\rangle \\
R &\to& \chi_R(w,r,n) D^{3/2}_{\frac{w}{2}\frac{r+1}{2}}|n\rangle
\end{eqnarray}
where $\chi_L(w,r,n)$ and $\chi_R(w,r,n)$ are the standard fermionic
chiral fields for the particle labelled $(w,r,n)$.

Then
\be
{\cal{M}}(w,r,n) = \langle n|\bar D^{3/2}_{\frac{w}{2}\frac{r+1}{2}}
D^{3/2}_{\frac{w}{2}\frac{r+1}{2}}|n\rangle
(\bar\chi_L\varphi\bar\chi_R + \bar\chi_R\bar\varphi\chi_L)
\ee
By the argument of the standard model
\be
\bar\chi_L\varphi\chi_R + \bar\chi_R\bar\varphi\chi_L
\ee
may be reduced to
\be
\rho(\bar\chi_L\chi_R + \bar\chi_R\chi_L) = \rho\bar\chi\chi
\ee
where $\rho$ is the vacuum expectation value of $\varphi$, the Higgs
field.  Then by (20.1)
\be
{\cal{M}}(w,r,n) = m(w,r,n)\bar\chi\chi
\ee
and by (20.4)
\be
m(w,r,n) = \rho(w,r)\langle n|\bar D^{3/2}_{\frac{w}{2}
\frac{r+1}{2}}D^{3/2}_{\frac{w}{2}\frac{r+1}{2}}|n\rangle
\ee
Then the four spectra (neutrinos, leptons, down quarks, up quarks)
may be expressed as follows:
\begin{subequations}
\begin{eqnarray}
m_\nu(n) &=& \rho(\nu)\langle n|b^3\cdot \bar b^3|n\rangle \\
m_\ell(n) &=& \rho(\ell)\langle n|\bar a^3\cdot a^3|n\rangle \\
m_d(n) &=& \rho(d)\langle n|\bar b^2\bar a\cdot ab^2|n\rangle \\
m_u(n) &=& \rho(u)\langle n|a^2b\cdot\bar b\bar a^2|n\rangle
\end{eqnarray}
\end{subequations}
where the four prefactors $(\rho(\nu),\rho(\ell),\rho(d),\rho(u))$
are intended to represent the products of 
the vacuum expectation value computed at the four 
local minima in the Higgs
potential with the numerical factors in $D^{3/2}_{mm^\prime}$.
The magnitude of $\rho$ sets the energy scale and differs for each
family.  The expressions (20.9) are based on $SU_q(2)$ as in Table
(11.1).

The spectrum of states allowed by the algebra is infinite but there
are only three particles in each family.  Without additional experimental input we have tentatively assigned these three
particles to the states $n=0,1,2$, in order of mass where $n=0$
corresponds to the lightest particle.
The masses in 
each spectrum are all proportional to the same $\rho$; and hence
the mass ratios may be computed without ambiguity in terms of the
two parameters ($q$ and $\beta$) of the model.  There are two
independent ratios that we choose as
\be
M = \frac{m(1)}{m(0)} \qquad \mbox{and} \qquad 
m = \frac{m(2)}{m(1)}
\ee
and that we may express as functions of $q$ and $\beta$.  By 
(20.19a)-(20.19d) one finds for the four families$^{9,11}$
%\begin{subequations}
\begin{eqnarray}
%\begin{array}{rl}
\mbox{neutrinos:} & & ~~~~~~M = m=q^6 \\
\mbox{leptons:} & &\frac{m-1}{m-q^6} = q^3\frac{M-1}{M-q^6}~, \quad
|\beta|^2 = q^6 \frac{M-1}{M-q^6} \\
\mbox{down quarks:} & &\frac{m-q^4}{m-q^6} = q^2\frac{M-q^4}{M-q^6}~, 
\quad
|\beta|^2 = \frac{m-q^4}{m-q^6} \\
\mbox{up quarks:} & &\frac{m-q^2}{m-q^6} = q^2\frac{M-q^2}{M-q^6}~, \quad
|\beta|^2 = \frac{M-q^2}{M-q^6} 
%\end{array}
\end{eqnarray}
%\end{subequations}

\vskip.3cm

The empirical input depends on the masses of the elementary fermions.
These are well determined for the leptons $(e,\nu,\tau)$, but for
the quarks they are not even well defined.  Since the quarks do not
exist as free particles, the quoted masses depend on the
theoretical procedure for defining them.  There is then a range of
``masses" given by the Particle Data Group.  Our treatment of mass is
limited by its dependence on the mass term of the standard theory,
as well as by an arbitrary assignment of $n$, and because the
binding associated with the gluon and gravitational fields is
either ignored or in some indirect way recognized in the mass term.

As already noted, the model permits higher excited states,
but no fourth generation
particles have been found, and the fourth generation lepton is
already excluded by the known width of the $Z^0$.  Additional
physical restrictions on the model are
therefore required.  It is possible, for example,
that these $q$-solitons will dissociate into preons at energies
below the mass of the fourth excited state.  The dissociation
energy would depend on the dynamics of the preons.

The neutrino mass spectrum is also a strong constraint on the model;
at present the data on this spectrum are compatible with $q\cong 1$.
In applications to fermionic currents, both in the lepton-neutrino
sector and in the Kobayashi-Maskawa sector, the data are compatible
with $q\cong 1$.

\section{Gluon Charge$^{15,17}$}

The previous considerations are based on electroweak physics.
To describe the strong interactions it is necessary according to
standard theory to introduce $SU(3)$ charge.  We shall 
therefore assume that
each of the four preon operators appears in triplicate
$(a_i,b_i,c_i,d_i)$ where $i=R,Y,G$, without changing the algebra
$(A)$. We shall assume that
these colored preon operators provide a basis for the
fundamental representation of $SU(3)$ just as the colored quark
operators do in standard theory, i.e. that color is not an emergent
property but appears already at the preon level.  

To adapt the electroweak operators
to the requirements of the standard theory we make the following
replacements:
%\be
\begin{eqnarray}
{\rm leptons}:  & &a^3 \to \epsilon^{ijk}a_ia_ja_k \\
{\rm neutrinos}:  & & c^3 \to \epsilon^{ijk}c_ic_jc_k \\
{\rm down~quarks}:  & & ab^2 \to a_i(\bar b^kb_k) \\
{\rm up~quarks}:  & & cd^2 \to c_i(\bar d^kd_k)
\end{eqnarray}
%\ee
where $\bar b^k$ and $\bar d^k\sim \bar 3$
representation of $SU(3)$ and $(i,j,k) =
(R,Y,G)$ and $(a_ib_ic_id_i)$ are creation operators for colored
preons.  
Then the creation operators for the
leptons and neutrinos are color singlets while
the creation operators for the quark states provide a basis for the  
fundamental representation of
$SU(3)$, as required by standard theory. 

Here $b$ and $\bar b$, as well as $d$ and $\bar d$, are antiparticles with respect to
$SU(3)$ but have the same values of $t_3$ and $t_0$.  Alternatively
replace $\bar b^kb_k$ and $\bar d^kd_k$ by
$g^{k\ell}b_kb_\ell$ and $g^{k\ell}d_kd_\ell$ respectively where
$g^{k\ell}$ is the group metric of $SU(3)$.

\vskip.5cm

\section{The Complementary Models$^{18}$}

We have ascribed to the quantum knot the state function
$D^j_{mm^\prime}(q|abcd)$, an irreducible representation of the
knot algebra $SL_q(2)$, where the indices $j=\frac{N}{2}$, $m=\frac{w}{2}$, $m^\prime = \frac{\pm r+1}{2}$ are restricted to values of $(N,w,r)$
allowed by the classical (geometrical) knots.  The quantum knots have
more degrees of freedom than their classical images with the
consequence that two quantum knots may be distinguishable when their
classical images are topologically indistinguishable.  

In particular there are four
distinguishable quantum trefoils with $(w,r) = (\pm 3,\pm 2)$ but only
two of their classical images $(w,r) = (\pm 3,2)$ are 
topologically different.  In
the physical application $(w,r) = (\pm 3,2)$ describe the leptons and
neutrinos while $(w,r) = (\pm 3,-2)$ describe the two varieties of
quarks, i.e., the two additional quantum knots are required to permit
the description of hypercharge and colored fermions.

These considerations have led us to two complementary models of the
elementary particles, namely

\ve

\begin{description}
\item{(a)} quantum knots
\item{(b)} preon structures
\end{description}
that are the field and particle descriptions of the same particles.
The correspondence may be expressed by the following relations
according to (14.9) and (14.10)
\begin{eqnarray}
w &=& n_a-n_d+n_b-n_c(=2m=-6t_3) \\
r+1 &=& n_a-n_d-n_b+n_c(=2m^\prime=-6t_0) \\
N &=& n_a+n_b+n_c+n_d(=2j=6t)
\end{eqnarray}
Here $(N,w,r)$ describe the number of crossings, the writhe and the
rotation of the particle regarded as a quantum knot of field while
$(n_a,n_b,n_c,n_d)$ record the number of $(a,b,c,d)$ preons in the
dual description of the same structure.  

We have also described this
particle by
\be
D^j_{mm^\prime} = D^{3t}_{-3t_3-3t_0} =
D^{N/2}_{\frac{w}{2}\frac{r+1}{2}}
\ee

The knot $(N,w,r)$ and the preon $(n_a,n_b,n_c,n_d)$ descriptions share
the same representation of $SL_q(2)$ as follows:

In terms of $(N,w,r)$ one has $D^j_{mm^\prime} = 
D^{N/2}_{\frac{w}{2}\frac{r+1}{2}}$ where
\be
D^{N/2}_{\frac{w}{2}\frac{r+1}{2}}(q|abcd) = \left[
\frac{\langle n_+^\prime\rangle!\langle n_-^\prime\rangle!}
{\langle n_+\rangle!\langle n_-\rangle!}\right]^{1/2}
\sum_{\scriptstyle 0\leq s\leq n_+\atop\scriptstyle 0\leq t\leq n_-}
\left\langle\matrix{n_+\cr s}\right\rangle_{q_1}
\left\langle\matrix{n_-\cr t}\right\rangle_{q_1}
\delta(s+t,n_+^\prime)a^sb^{n_+-s}c^td^{n_--t}
\ee
and again in terms of $(N,w,r)$
\begin{eqnarray}
n_\pm &=& \frac{1}{2}[N\pm w] \\
n_\pm^\prime &=& \frac{1}{2}[N\pm(r+1)]
\end{eqnarray}

The complementary description expressed
in terms of the population numbers $(n_a,n_b,n_c,n_d)$ is
\be
D^j_{mm^\prime} = {\cal{D}}^{N^\prime}_{\nu_a\nu_b}
\ee
where $N^\prime$ is by (14.8) the total number of preons
\be
{\cal{D}}^{N^\prime}_{\nu_a\nu_b} = \left[\frac{\langle n_a+n_c\rangle!
\langle n_b+n_d\rangle!}{\langle n_a+n_b\rangle!
\langle n_c+n_d\rangle!}\right]^{1/2} 
\sum_{\scriptstyle N^\prime\geq n_a,n_b\geq 0\atop\scriptstyle
N^\prime\geq n_c,n_d\geq 0}
\left\langle\matrix{n_a+n_b\cr n_a}\right\rangle_{q_1}
\left\langle\matrix{n_c+n_d\cr n_c}\right\rangle_{q_1}
a^{n_a}b^{n_b}c^{n_c}d^{n_d}
\ee
The limits on $\sum$, literally translated from $D^j_{mm^\prime}$ are shown in the expression for ${\cal{D}}^{N^\prime}_{\nu_a\nu_b}$ but these
limits simply describe the requirement that all population numbers,
$n_i$ satisfy $N^\prime \geq n_i \geq 0$, since $N\geq w \geq 0$.

Here the exponents $(n_a,n_b,n_c,n_d)$ of $(abcd)$ are 
\be
\begin{array}{ll}
n_a = s & n_b = n_+-s \\
n_c = t & n_d = n_--t
\end{array}
\ee
They are related to $(n_+,n_-,n^\prime_+,n_-^\prime)$ by
\be
\begin{array}{ll}
n_+ = n_a+n_b & n_+^\prime = n_a+n_c \\
n_- = n_c+n_d & n_-^\prime = n_b+n_d
\end{array}
\ee
Since $a$ and $d$ have opposite charge and hypercharge, while $b$ and
$c$ are neutral with opposite hypercharge, we may define the ``preon
numbers" $\nu_a$ and $\nu_b$ as follows
\be
\begin{array}{rcl}
\nu_a &=& n_a-n_d \\
\nu_b &=& n_b-n_c
\end{array}
\ee
The preon numbers are the same for every term of 
${\cal{D}}^{N^\prime}_{\nu_a\nu_b}$,
in (22.9), since
\be
\begin{array}{rcl}
\nu_a + \nu_b &=& 2m=w \\
\nu_a-\nu_b &=& 2m^\prime = r+1
\end{array}
\ee
By (22.13)
the conservation of the writhe and rotation is equivalent to 
 the conservation of the preon numbers $\nu_a$ and $\nu_b$, and the
kinematic factor is described equally well by $(N,w,r)$ and
$(N,\nu_a,\nu_b)$.  Viewed as twisted loops, the preons 
could be prevented
from unrolling into simple loops by the 
dynamical conservation of writhe and
rotation (computed in the same way as for knots).  Viewed as a
particle the preon 
could be conserved by the dynamical conservation of preon numbers.

The trefoil solutions of the equations (22.1)-(22.3)
relating $(N,w,r)$ to
$(n_a,n_b,n_c,n_d)$ are given in Table 22.1:
\ve

\ce{\bf Table 22.1}

\be
\begin{array}{ccccc}
\quad &  \underline{n_a} \quad & \underline{n_b} \quad & \underline{n_c} \quad & \underline{n_d}  \\ \nonumber
\ell \quad & 3 \quad & 0  \quad & 0 \quad & 0  \\ \nonumber
\nu \quad & 0 \quad & 0 \quad & 3 \quad & 0 \\
d \quad & 1 \quad & 2 \quad & 0 \quad & 0 \\ \nonumber
u \quad & 0 \quad & 0 \quad & 1 \quad & 2 
\end{array}
\ee

\vskip.3cm

In general
\be
{\cal{D}}^{N^\prime}_{\nu_a\nu_b} = \sum_{Nwr} \delta(N^\prime,N) 
\delta(\nu_a+\nu_b,w)
\delta(\nu_a-\nu_b,r+1) D^{N/2}_{\frac{w}{2}\frac{r+1}{2}}
\ee

Since the number of crossings equals the number of preons, one may
speculate that there is one preon at each crossing if both preons and
crossings are considered pointlike. 
If the pointlike crossings are
labelled $(\vec x_1\vec x_2\vec x_3)$, then the wave
functions of the trefoils representing leptons $(\ell)$, neutrinos
$(\nu)$, down quarks $(d)$, up quarks $(u)$ are as follows:
\begin{eqnarray}
\Psi_\ell(\vec x_1\vec x_2\vec x_3) &=& \epsilon^{ijk}\psi_i
(a|\vec x_1)\psi_j(a|\vec x_2)\psi_k(a|\vec x_3) \\ 
\Psi_\nu(\vec x_1\vec x_2\vec x_3) &=& \epsilon^{ijk}
\psi_i(c|\vec x_1)\psi_j(c|\vec x_2)\psi_k(c|\vec x_3) \\
\Psi_d(\vec x_1\vec x_2\vec x_3) &=& \psi_i(a|\vec x_1)
\bar\psi^j(b|\vec x_2)\psi_j(b|\vec x_3) \\
\Psi_u(\vec x_1\vec x_2\vec x_3) &=& \psi_i(c|\vec x_1)
\bar\psi^j( d|\vec x_2)\psi_j(d|\vec x_3)
\end{eqnarray}
where $i=(R,Y,G)$ and $\psi_i(a|\vec x) \ldots \psi_i(d|\vec x)$
are colored $\delta$-like functions localizing the preons at the
crossings.

Then the wave function of a lepton describes a \underline{singlet} 
trefoil
particle containing three preons of charge $(-e/3)$ and
hypercharge $(-e/6)$.  

The corresponding characterization of a
neutrino describes a \underline{singlet} trefoil 
containing three neutral
preons of hypercharge $(-e/6)$.

The wave function of a down quark describes a colored trefoil particle
containing one $a$-preon with charge $(-e/3)$ and hypercharge
$(-e/6)$ and two neutral $b$-preons with hypercharge $(e/6)$.  

The
corresponding characterization of an up-quark describes a colored
trefoil containing two charged $d$-preons with charges $(e/3)$
and hypercharge $(e/6)$, and one neutral $c$-preon with hypercharge
$(-e/6)$.

This hypothetical structure is held together by the trefoil of field
connecting the charged preons.  A search for this kind of substructure
depends critically on the mass of the conjectured preons and the
strength with which they are bound.  Since there is no empirical
information to guide us in discussing the hypothetical preons,
either in fixing the masses of the fermionic preons or in
determining the fields comprising the binding trefoil, we shall assume
that the preonic fermions and bosons conform to the same general rules
as the familiar fermions and bosons.  Under these assumptions we shall
now consider the masses of the fermionic preons and the interactions
of the bosonic preons.

\vskip.5cm

\section{Mass of Preons}

Let us assume that the mass of the preon is computed in
the same way as we have computed
the mass of the elementary fermions, i.e., by 
adopting the mass terms of the standard theory, namely$^1$
\be
{\cal{M}} = \bar L\varphi R + \bar R\varphi L
\ee
where $L$ and $R$ are left and right chiral spinors and $\varphi$ is
the Higgs scalar.

We shall assign a $SU_q(2)$ singlet structure to $\varphi$ 
(unitary gauge) and the 
preon representation $D^{1/2}_{mm^\prime}$ to both $L$ and $R$.  Then
we substitute for $L$ and $R$ as follows:
\begin{eqnarray}
L \to \chi_L D^{1/2}_{mm^\prime}|0\rangle \\
R \to \chi_R D^{1/2}_{mm^\prime}|0\rangle
\end{eqnarray}
where $\chi_L$ and $\chi_R$ are standard fermionic fields and
$D^{1/2}_{mm^\prime}|0\rangle$ describes the internal structure of 
the preons.  Here $|0\rangle$ is the ground state of the $SU_q(2)$
algebra.  Then
\begin{eqnarray}
{\cal{M}} &\equiv& \langle 0|\bar D^{1/2}_{mm^\prime}D^{1/2}_{mm^\prime}
|0\rangle(\bar\chi_L\varphi\chi_R + \bar\chi_R\varphi\chi_L) \\
& &\mbox{}=M(m,m^\prime)\bar\chi\chi
\end{eqnarray}
where the mass is
\be
M(m,m^\prime) = \rho(m,m^\prime)\langle 0|\bar D^{1/2}_{mm^\prime}
D^{1/2}_{mm^\prime}|0\rangle
\ee
where $\rho(m,m^\prime)$ is the vacuum expectation value of the Higgs
field at
a local minimum of the ``Higgs potential". 

We
shall assume that there are 4 local minima of the Higgs potential, 
so that
\be
\rho(m,m^\prime) = \rho\left(\pm\frac{1}{2},\pm\frac{1}{2}\right)
\ee
For example, the mass of the $D^{1/2}_{\frac{1}{2}\frac{1}{2}}$ preon
is
\be
M\left(\frac{1}{2},\frac{1}{2}\right) = \rho\left(\frac{1}{2},
\frac{1}{2}\right) \langle 0|\bar aa|0\rangle
\ee
The mass of the electron computed in the same way is
\be
M\left(\frac{3}{2},\frac{3}{2}\right) = 
\rho\left(\frac{3}{2},\frac{3}{2}\right) \langle 0|\bar a^3a^3|0\rangle
\ee
Then the ratio of the preon mass $(m_p)$ to the electron mass
$(m_e)$ is
\be
\frac{m_p}{m_e} = \frac{\rho\left(\frac{1}{2},\frac{1}{2}\right)}
{\rho\left(\frac{3}{2},\frac{3}{2}\right)}
\frac{\langle 0|\bar aa|0\rangle}{\langle 0|\bar a^3a^3|0\rangle}
\ee

The model leads to additional
relations between the corresponding Higgs fields.

Let the vacuum expectation values of these fields be 
$(\rho_a,\rho_b,\rho_c,\rho_d)$ where
\be
(\rho_a,\rho_b,\rho_c,\rho_d) = \left(\rho\left(\frac{1}{2}\frac{1}{2}
\right),~\rho\left(\frac{1}{2},-\frac{1}{2}\right),~\rho\left(
-\frac{1}{2},\frac{1}{2}\right),~\rho\left(-\frac{1}{2},-\frac{1}{2}
\right)\right)
\ee
respectively.  Then under the assumption that the quantum group is
$SU_q(2)$, one has
\be
\begin{array}{rcl}
d &=& \bar a \\ c &=& -q_1\bar b
\end{array}
\ee
and the mass ratio of $a$ to $d$ is

\begin{eqnarray}
\frac{m_a}{m_d} &=& \frac{\rho_a}{\rho_d}
\frac{\langle 0|\bar aa|0\rangle}{\langle 0|\bar dd|0\rangle} \\
&=& \frac{\rho_a}{\rho_d}\frac{1-q_1^2|\beta|^2}
{1-|\beta|^2}
\end{eqnarray}
Similarly the mass ratio of $b$ to $c$ is
\begin{eqnarray}
\frac{m_b}{m_c} &=& \frac{\rho_b}{\rho_c}
\frac{\langle 0|\bar bb|0\rangle}{\langle 0|\bar cc|0\rangle} \\
&=& q^2 \frac{\rho_b}{\rho_c}
\end{eqnarray}
Since $a$ and $d$ and also $b$ and $c$ are antiparticles, their masses
are equal under the usual forms of the quantum theory that conserve 
TCP.

Then within the limitations of the model
\begin{eqnarray}
\frac{\rho_a}{\rho_d} &=& \frac{1-|\beta|^2}{1-q_1^2|\beta|^2} \\
\frac{\rho_b}{\rho_c} &=& q_1^2
\end{eqnarray}

\vskip.5cm

\section{Electroweak Interactions of Preons}

The weak vectors of standard theory belong to the $j=3$ and the
$j=0$ representations of the $SU_q(2)$ quantum group.  Of these only
the $j=0$ vector interacts with the $j=1/2$ preons.  The $j=1$
vector preon will, however, connect with the $j=1/2$ states.  This
is a new interaction 
that would contribute to the binding of the preons into
a composite particle.

If the internal states are consistently represented by 
$D^{3t}_{-3t_3-3t_0}$ the weak vector bosons 
$(\vec W)$ of the standard theory
correspond to $j=3$ and $t=1$ as previously shown and given in
Table 24.1:

\ce{\bf Table 24.1}
\vskip.3cm

\[
\begin{array}{ccccl} \hline
& t & t_3 & t_0 & D^{3t}_{-3t_3-3t_0} \\ \hline
W^+ & 1 & 1 & 0 & D^3_{-30} \sim\bar b^3\bar a^3 \equiv
{\cal{D}}_+(1) \\
W^- & 1 & -1 & 0 & D^3_{30}\sim a^3b^3\equiv {\cal{D}}_-(1) \\
W^3 & 1 & 0 & 0 & D^3_{00}\sim f_3(\bar bb)\equiv 
{\cal{D}}_0(1) \\ \hline
\end{array}
\]

\no The corresponding states of the vector preons with $j=1$ and 
$t=1/3$ are given in Table 24.2:

\ce{\bf Table 24.2}
\vskip.3cm

\[
\begin{array}{ccccl} \hline
& t & t_3 & t_0 & D^{3t}_{-3t_3-3t_0} \\ \hline
W^+ & \frac{1}{3} & \frac{1}{3} & 0 & D^1_{-10} = cd \sim \bar b\bar a
\equiv {\cal{D}}_+(1/3) \\
W^- & \frac{1}{3} & -\frac{1}{3} & 0 & D^1_{10} = ab \equiv
{\cal{D}}_-(1/3) \\
W^3 & \frac{1}{3} & 0 & 0 & D^1_{00} = ad+bc \equiv 1-(1+q_1)
\bar bb = {\cal{D}}_0(1/3) \\ \hline
\end{array}
\]
The elements of $D^{3t}_{-3t_3-3t_0}$ representing the vector triplets
in both tables may be read as composite creation operators with
$a,b,c,d$ carrying the correct charge and hypercharge for fermionic
preons.  In both cases the operators ${\cal{D}}_i(t)$ satisfy the
following commutation relations
\begin{eqnarray}
& &[{\cal{D}}_i(t),{\cal{D}}_j(t)] = c_{ij}^k(t|b\bar b)
{\cal{D}}_k(t) \\
& & t = 1,\frac{1}{3} \quad \mbox{and} \quad (i,j,k) =
(+,-,3) \nonumber
\end{eqnarray}
We introduce the covariant derivatives
\be
\nabla_\mu(t) = 1~\partial_\mu + {\cal{W}}_\mu(t)
\ee
with the matrix vector potential
\be
{\cal{W}}_\mu(t) = W^-(t)t_-{\cal{D}}_-(t) +
W^+(t)t_+{\cal{D}}_+(t) + W^3(t)t_3{\cal{D}}_3(t)
\ee
where
\be
t_+ = \left(
\begin{array}{cc}
0 & 0 \\ 1 & 0
\end{array} \right)~, \quad t_- = \left(
\begin{array}{cc}
0 & 1 \\ 0 & 0
\end{array} \right)~, \quad t_3 = \left(
\begin{array}{cc}
1 & 0 \\ 0 & -1
\end{array} \right)
\ee
and
\be
[t_i,t_j] = c_{ij}^kt_k
\ee
In the preceding equations (24.3)-(24.5) we have here adopted the
same form for the weak preon vectors as for the standard weak
vectors.  The vector field strengths are
\be
{\cal{W}}_{\mu\lambda}(t) = [\nabla_\mu(t),\nabla_\lambda(t)]
\ee
and the electroweak field lagrangian contains the following invariants:
\be
L(t) = {\rm Tr}\langle 0|{\cal{W}}_{\mu\lambda}(t)
{\cal{W}}^{\mu\lambda}(t)|0\rangle \qquad t = 1,\frac{1}{3}
\ee
The preon interactions are mediated by the vector fields for which
$t=1/3$ and not by the vector fields of the standard theory for
which $t=1$.  These interactions are induced by the kinetic term:
\be
\langle 0|\bar P\nabla\!\!\!\!/\left(\frac{1}{3}\right)P|0\rangle
\ee
where $P$ is the doublet 
$\left(
\begin{array}{c}
a \\ c
\end{array} \right)
$
and $\nabla\!\!\!\!/\left(\frac{1}{3}\right)$
is given by (24.2).

The field lagrangian, $L(t)$, reduces to a form similar to that of
a lagrangian of a non-Abelian vector field that is derived from a
Lie algebra.  Unlike the standard case, however, $L(t)$ has structure
constants that depend on $q$ and $\beta$. 

\vskip.5cm

\section{Remarks$^{18}$}

The model described in this review is constructed by replacing the
point particles of the standard model by quantum knots, described as
members of the irreducible representations of $SL_q(2)$, the knot
algebra.  This replacement is carried out by multiplying the normal
modes and hence the field operators of the standard model by the state functions of the quantum knots.  The essential speculative 
elements of the modified standard
model include (a) the Higgs potential, necessary for
standard electroweak theory, and (b) the quantum group $SL_q(2)$
essential for the conceptual simplification of the electroweak theory
that is described here.  Both the Higgs potential and $SL_q(2)$, raise unanswered questions.  Let us consider $SL_q(2)$ first.

The gauge group of the $SL_q(2)$ algebra permits a classification
of the elementary fermions and hypothetical preons as matrix elements
of the $j=3/2$ and $j=1/2$ representations respectively of either
$SL_q(2)$ or $SU_q(2)$.  Since these quantum groups describe familiar
symmetries when $q=1$, and since only the deviation of $q$ from
unity permits the conceptually simpler picture that the quantum groups
permit, the view that one takes of the parameter,
$q$, becomes important for the view that one takes of our use of
$SL_q(2)$ or $SU_q(2)$ itself.  Like Planck's constant, which 
normalizes the non-Abelian Heisenberg algebra, the parameter $q$ also
normalizes a non-Abelian algebra but 
an algebra dependent on $\epsilon_q$ instead
of $i$, where $\epsilon_q$ is also a square root of -1.  Unlike
$\hbar$ which has the dimension of an action, the constant $q$ is
dimensionless. 

The introduction of substructure, determined by the $SL_q(2)$
algebra, for the quantum fields in terms of preons resembles the
introduction of substructure for the classical fields in terms of
field quanta determined by the Heisenberg algebra holding for conjugate
field operators.  This analogy suggests a comparison of the
$SL_q(2)$ algebra determined by $q$ with the Heisenberg algebra
determined by $h$.

The $SL_q(2)$ algebra leaves invariant the quadratic form:
\be
K = A^t\epsilon_q A
\ee
under the transformations
\be
A^\prime = TA \qquad T\epsilon SL_q(2)
\ee
where
\be
\epsilon_q = \left(
\begin{array}{cc}
0 & q^{-/2} \\ -q^{1/2} & 0
\end{array} \right)
\ee
as defined in (3.3).

Let us normalize the invariant $K$ by setting $K=q^{-1/2}$.  If we
now choose
\be
A = \left(
\begin{array}{c}
D_x \\ x
\end{array} \right)
\ee
we have by (25.1) the $SL_q(2)$ invariant relation
\be
D_xx - qxD_x = 1
\ee
This equation may be satisfied if $D_x$ is chosen as the $q$-difference
operator:
\be
D_x\psi(x) \equiv \frac{\psi(qx)-\psi(x)}{qx-x}
\ee
Let
\be
P_x = \frac{\hbar}{i} D_x
\ee
Then
\be
(P_xx-qxP_x) \psi(x) = \frac{\hbar}{i} \psi(x)
\ee
If $q$ is near unity
\be
q = 1+\delta
\ee
then
\be
D_x\psi(x) = \frac{\psi(x+\delta x)-\psi(x)}{\delta x}
\ee
In the limit $\delta=0$
\be
D_x\psi(x) = \frac{d}{dx} \psi(x)
\ee
and
\be
P = \frac{\hbar}{i} \frac{d}{dx}
\ee
Then (25.8) becomes the Heisenberg commutator
\be
(P_xx-xP_x)\psi(x) = \frac{\hbar}{i} \psi(x)
\ee
The operator $D_x$ may also be expressed in the notation of
``basic numbers", which are useful for discussing $SL_q(2)$, as
follows:  Let
\be
\theta = x\frac{d}{dx}
\ee
Then
\be
q^\theta f(x) = f(qx)
\ee
and by (25.6)
\be
xD_x = \frac{q^\theta-1}{q-1}
\ee
or
\begin{eqnarray}
xD_x &=& \langle\theta\rangle \\
&=& \langle x\frac{d}{dx}\rangle
\end{eqnarray}
so that $xD_x$ is a ``basic dilatation operator".  If $q$ is near
unity, $D$ resembles the differentiation operator on a lattice space
and $q$ may play the role of a dimensionless regulator.

In view of the physical evidence suggestive of substructure that has
been described here, as well as the natural appearance of the
non-standard $q$-derivatives, it may be possible to utilize
$SL_q(2)$ to describe a finer level of structure than is currently
considered.

On the other hand, the basic question is whether $SL_q(2)$ or $SU_q(2)$
are fundamental symmetries, and whether $q$ is accordingly a
fundamental physical constant that comes out differently in
different contexts, depending on other neglected physics; or whether
the $q$-model is simply an effective theory, where $q$ and $\beta$
are surrogates for physics ignored in the standard model.

Let us finally consider the Higgs potential which has a single
minimum in the standard model.  The calculationn of mass described
here is not an essential part of the knot model but presents an
interesting possibility if one assumes that the ``Higgs potential"
has a sequence of local minima associated with one or more scalar
fields.  Then by consistent use of the Higgs mass term it is in
principle possible to calculate all the masses of the model in terms
of these minima.  The question then would be how these minima should
be regarded if they exist.  They might be considered simply as given
data determining boundary conditions on the model.  On the other
hand, since the gravitational field has been ignored and since all
masses are sources of the gravitational field, there is the
possibility that the complete set of ``Higgs scalars" might appear as
part of the expanded Einstein field, especially since scalar fields
appear in supergravity theories.

\vskip 0.2in
\noindent
{\bf Acknowledgement:}  I thank J. Smit and A.C. Cadavid for helpful discussions.
\ve

\no {\bf References}

\begin{enumerate}
\item Thomson, W. H., Trans. R. Soc. Edinb. {\bf 25}, 217-220 (1969).
\item Faddeev, L. and Niemi, Antti, J., Nature {\bf 387}, May 1 (1997).
\item Kauffman, L. H., Int. J. Mod. Phys. A{\bf 5}, 93 (1990).
\item {\it Quantum Groups}, ed. by T. Curtright, D. Fairlee,
C. Zachos, World Scientific (1991).
\item Reshetikhin, N. Yu., Takhtadzhyan, L. A., and Faddeev, L. D.,
Leningrad Math J. {\bf 1} (1990).
\item Woronowicz, J., Comm. Math. Phys. {\bf 111}, 613 (1987).
\item Gasper, G. and Rahman, M., {\bf Hypergeometric Series},
Cambridge University Press (1990).
\item Weyl, H., {\it Theory of Groups and Quantum Mechanics}, ed.
E. P. Dutton (1931).
\item Finkelstein, R. J., J. Math. Phys. {\bf 8}, 443 (1967).
\item Finkelstein, R. J., Int. J. Mod. Phys. A{\bf 20}, 6487 (2005).
\item Cadavid, A. C. and Finkelstein, R. J., Int. J. Mod. Phys.
A{\bf 21}, 4269 (2006).
\item Finkelstein, R. J., Int. J. Mod. Phys. A{\bf 22}, 4467 (2007).
\item Groza, V. A., Kacharik I., and Klimyk, A. U., J. Math.
Phys. {\bf 31}, 2769 (1990).
\item Cadavid, A. C. and Finkelstein, R. J., J. Math. Phys. {\bf 36},
1912 (1995).
\item Finkelstein, R. J., Int. J. Mod. Phys. A{\bf 24}, 2307 (2009).
\item arXiv-hep/th 07011 24.
\item arXiv-hep/th 0901 1687.
\item arXiv-hep/th 0912 3552.
\item arXiv-hep/th 1011 0764.
\end{enumerate}

\end{document}